%% file: Mauscript_AFM.tex
\documentclass[reprint,onecolumn,amsmath,amssymb,jap,aip,superscriptaddress]{revtex4-2}
\usepackage{graphicx}
\usepackage{dcolumn}
\usepackage{bm}
\usepackage{epsfig}
\usepackage{mathptmx}
\usepackage{times}
\usepackage{dcolumn}
\usepackage{units}
\usepackage{upgreek}
\usepackage{tabularx}
\usepackage{threeparttable}
\usepackage[space]{grffile}
\usepackage[normalem]{ulem}

\usepackage{subfig}
\usepackage{multirow}
\usepackage{array}
\usepackage{xcolor}
\usepackage{soul}
\usepackage[]{caption}
\usepackage{physics}
\usepackage[T1]{fontenc}
\usepackage{units}
\usepackage{tgschola}

\newcommand{\GST}{\textit{Ge$_2$Sb$_2$Te$_5$}}
\newcommand{\RESET}{\textit{RESET}}
\newcommand{\SET}{\textit{SET}}
\newcommand{\figref}[2]{Figure~\ref{#1}#2}

\begin{document}

\title{Cryogenic In-Memory Computing with Phase-Change Memory}

\author{Davide G. F. Lombardo}
\affiliation{IBM Research -- Europe, S\"{a}umerstrasse 4, 8803 R\"{u}schlikon, Switzerland}
\affiliation{École polytechnique fédérale de Lausanne (EPFL), Lausanne, Switzerland}
\author{Siddharth Gautam}
\affiliation{IBM Research -- Europe, S\"{a}umerstrasse 4, 8803 R\"{u}schlikon, Switzerland}
\affiliation{École polytechnique fédérale de Lausanne (EPFL), Lausanne, Switzerland}
\author{Alberto Ferraris}
\affiliation{IBM Research -- Europe, S\"{a}umerstrasse 4, 8803 R\"{u}schlikon, Switzerland}
\affiliation{École polytechnique fédérale de Lausanne (EPFL), Lausanne, Switzerland}
\author{Manuel Le Gallo}
\affiliation{IBM Research -- Europe, S\"{a}umerstrasse 4, 8803 R\"{u}schlikon, Switzerland}
\author{Abu Sebastian}
\affiliation{IBM Research -- Europe, S\"{a}umerstrasse 4, 8803 R\"{u}schlikon, Switzerland}
\author{Ghazi Sarwat Syed}
\affiliation{IBM Research -- Europe, S\"{a}umerstrasse 4, 8803 R\"{u}schlikon, Switzerland}
\email{ghs@zurich.ibm.com, ase@zurich.ibm.com}

\maketitle

\section*{Abstract}

\noindent{In-memory computing (IMC) is an emerging non-von Neumann paradigm that leverages the intrinsic physics of memory devices to perform computations directly within the memory array. Among the various candidates, phase-change memory (PCM) has emerged as a leading non-volatile technology, showing significant promise for IMC, particularly in deep learning acceleration. PCM-based IMC is also poised to play a pivotal role in cryogenic applications, including quantum computing and deep space electronics. In this work, we present a comprehensive characterization of PCM devices across temperatures down to \unit[5]{K}, covering the range most relevant to these domains. We systematically investigate key physical mechanisms such as phase transitions and threshold switching that govern device programming at low temperatures. In addition, we study attributes including electrical transport, structural relaxation, and read noise, which critically affect readout behavior and, in turn, the precision achievable in computational tasks.}

\begin{flushleft}
    \textbf{Keywords}: Phase Change Memory, Cryo-electronics, In-Memory Computing 
\end{flushleft}

\noindent Phase-change memory stores data by exploiting the large resistivity contrast between the amorphous and crystalline states of phase-change materials. Already a commercially established non-volatile memory technology, PCM has found use in both embedded and stand-alone products~\cite{GSSyedASebastian202506,MBoniardiARedaelli2025}. An emerging application of PCM is in-memory computing, a neuromorphic approach that exploits PCM device physics and crossbar arrays to perform computation within memory~\cite{MLeGalloASebastian202309,SAmbrogioGWBurr202308,GSSyedASebastian202312}.  A typical PCM device for these applications consists of a nanoscale volume of phase-change material sandwiched between two electrodes, programmable to different resistance states. The programming is largely understood to involve electro-thermal mechanisms driven by Joule heating~\cite{MLeGalloASebastian201601}. The phase transition from crystalline to amorphous (\RESET{} operation) requires melting and quenching the phase-change material, while the transition from amorphous to crystalline (\SET{} operation) occurs via crystal growth. The current-voltage characteristics governing read-out behavior of the devices are also thermally assisted since the bulk conductance mechanisms are strongly dependent on atomic disorder~\cite{DIelminiYZhang200709,DKrebsMWuttig2010,MLGalloDKrebs201509,MNardoneVGKarpov201210}. \\

More recently, PCM is being considered for non-volatile electronic storage in ultra-low temperature applications~\cite{JHQuintinoPalharesPGaly202409}. These include use in superconducting circuits~\cite{SAlamAAziz202303} and spacecraft~\cite{AKeysEKolawa2007}, the latter of which also benefits from the technology's intrinsic radiation hardness~\cite{APFerreiraMYousif201004}. There is also an emerging application in cryogenic IMC~\cite{YLiuQShao202504}. While temperature effects at and above room temperature are reasonably well understood, leading to established PCM device models in this range, there is a significant lack of data on device behavior at cryogenic temperatures. Notably, most studies~\cite{BKerstingMSalinga2019,BKerstingASebastian2021} focus on operation at high ambient temperatures since this is of interest for applications such as the automotive sector~\cite{PCappellettiPZuliani202003}. There is limited understanding of temperature effects below \unit[77]{K}, and measurements near \unit[5]{K} are entirely missing. Only recently has ultra-low temperature operation been studied~\cite{ABMHTalukderHSilva202402,JHQuintinoPalharesPGaly202409}; however, there is still very limited focus and few physical insights provided on the temperature dependence of the programming and read-out behaviors particularly, in the context of IMC.\\

In this work, we bridge the knowledge gap regarding the temperature dependence of the device by systematically examining its programming and read-out characteristics across a temperature range of \unit[5--400]{K}, under different phase configurations (amorphous volume fractions). The study is divided into three parts. The first part examines how key programming metrics scale under cryogenic conditions. This includes understanding the temperature dependence of programming efficiency during \RESET{} and threshold switching in \SET{} operations. The second part investigates the read-out characteristics, elucidating how different charge transport mechanisms influence resistance values, drift, and read noise of different phase configurations across the temperature span.  Finally, the third part discusses how these metrics ultimately impact IMC performance under cryogenic operating conditions. 

\section*{PCM programming at cryogenic temperatures}

\noindent Electro-thermal mechanisms are understood to govern the switching behavior of PCM devices~\cite{MLeGalloASebastian201601}. Briefly, the temperature rise in an active region of a device due to electrical power dissipation can be expressed by \({T_{\text{a}} = T_{\text{amb}} + R_{\text{th}} \cdot P_{\text{prog}}}\).  Here, $T_{\text{amb}}$ is the ambient temperature, $R_{\text{th}}$ is the effective thermal resistance to heat dissipation, and $P_{\text{prog}}$ is the power dissipated in the device during programming. When $T_{\text{a}}$ exceeds the melting temperature of the phase-change material, a portion of the material melts and can be \RESET{}. Similarly, in the amorphous state, switching is widely attributed to Joule heating: the temperature rise activates carriers, increasing conductivity and triggering a positive feedback loop of further heating, which ultimately leads to crystallization and the \SET{} operation. In the following we study the temperature dependence of these switching mechanisms. 

\subsubsection*{\RESET{} Operation}

\noindent \figref{fig2-prog}{a} presents the programming characteristics of a device measured across the full \unit[5$-$400]{K} ambient temperature span. In this experiment, \RESET{} pulses with increasing amplitude $V_\text{RESET}$ are applied to the device to form increasingly larger amorphous volumes. The programming current and voltage drops across the device are computed using the captured waveforms.  To ensure consistent programmability, the device is \SET{} between each \RESET{} pulse. This procedure is repeated five to ten times at each temperature. As one would expect, the programming current in the device increases as the ambient temperature decreases. That is, to form an amorphous volume of the same size via melt-quenching, more power must be dissipated at lower operating temperatures. This trend is further illustrated in \figref{fig2-prog}{b}, where we graph the programming power at which the device resistance begins to increase. This marks the point at which the active temperature $T_\text{a}$ reaches the melting temperature $T_\text{melt}$, against the corresponding ambient temperatures at which the measurements were conducted.  \\

As the ambient temperature decreases, more power is required to reach $T_\text{melt}$. The slope of plot in Figure \ref{fig2-prog}b corresponds to the effective thermal resistance, $R_{\text{th}}$, which is a direct measurement of thermal confinement in the device. By extrapolating the line to \unit[0]{$\mu$W}, we estimate the melting temperature. We find that data points above \unit[150]{K} can be reasonably fitted to $T_\text{a}$, providing a $T_\text{melt}$ of \unit[627]{$^\circ$C}, which is typical of GST.  The $R_\text{th}$ we extract is on the order of $1\,\text{K\,\textmu W}^{-1}$ that is typical of GST based mushroom-type devices~\cite{ASebastianDKrebs201407,MLGalloASebastian202003}. However, interestingly, we note that at cryogenic temperatures (data points \unit[75]{K} and \unit[5]{K}) the measured programming power is lower than would be predicted by the linear fit. We can explain this discrepancy by recognizing that the thermal resistance, $R_\text{th}$, is itself temperature dependent. As a first-order approximation, $R_\text{th}$ is inversely proportional to the thermal conductivity $\kappa$. And $\kappa$ is known to decrease with temperature in chalcogenide materials~\cite{YMGalperinVIKozub198901,KSSiegertMWuttig2015}. In a PCM device, heat is dissipated mostly through surrounding insulating layers, which also exhibit similar temperature dependence on $\kappa$. As a result, it is expected that $R_\text{th}$ must increase as $T_\text{amb} \rightarrow 0$, meaning that comparatively less power than predicted by the fit needs to be dissipated to reach the melt temperature.  In practice, this discrepancy is favorable, as the programming power does not increase uncontrollably with decreasing temperature, enabling peripheral circuits to reliably support device operation even under cryogenic conditions. \\

The second observation made in \figref{fig2-prog}{a} is that the memory window, defined as the ratio between the maximum \RESET{} and minimum achievable \SET{} resistance, widens with decreasing temperature (see \figref{fig2-prog}{c}). Notably, the SET/RESET increases from $2\,\text{k}\Omega / 5 \times 10^3\,\text{k}\Omega$ at $300\,\text{K}$  to $5\,\text{k}\Omega / 10^{10}\,\text{k}\Omega$ at $5\,\text{K}$. This represents an unprecedented resistance contrast exceeding $10^9$, only limited by the measurement unit.  This is again a favorable outcome arising from cryogenic conditions. A larger memory window can support an increased number of analog or multilevel states, which is beneficial for both memory and computing applications. 

\subsubsection*{\SET{} Operation}

\noindent We now discuss the temperature-dependent data gathered on threshold switching. In this measurement, the device was fully \RESET{} five to ten times at each temperature, and \SET{} traces were collected, from which the voltage drop across the device and the device current were computed. As shown in Figure \ref{fig2b-thresh}a, the threshold voltage exhibits a negative temperature dependence, increasing by nearly a factor of two as the temperature decreases from \unit[300]{K} to \unit[5]{K}. This critical dependence has been previously observed at temperatures exceeding room temperature and can be tied to the thermally activated nature of the carriers required in the threshold switching process. The observation that this characteristic persists even at very low temperatures supports the viability of \SET{} operation through threshld switching under cryogenic conditions. Conversely, we find that the threshold current (see Figure \ref{fig2b-thresh}b) decreases non-linearly with decreasing temperature, enabling threshold switching to be triggered at sub-$\mu$A levels at \unit[5]{K}. Threshold current is defined as the current flowing through the device in the moments prior to threshold switching. \\

Just like thermal conductivity, the specific heat capacity ($C$) also exhibits a strong dependence on temperature. Below the Debye temperature, $C$ begins to decrease rapidly below the Dulong-Petit limit. As a result, even a small amount of electrical current can lead to a rapid temperature increase, initiating a positive thermal feedback loop in the device~\cite{SSlesazeckTMikolajick2015}. In the transient (dynamic) regime, the heat equation governing the rate of temperature change, given by $\partial_t T \propto C_\mathrm{th}^{-1}$, is inversely proportional to the total thermal capacitance of the device, where $C_\text{th} \approx C \cdot V$ and $V$ is the active volume. Therefore, a lower heat capacity results in a steeper temperature rise under a given power input. This is further verified in Figure \ref{fig2b-thresh}c, where we plot the power required to trigger threshold switching, computed as the product of threshold voltage and current. Threshold power decreases as the temperature is lowered.\\

In summary, the following observations are made regarding the programming characteristics. For the \RESET{} operation, the programming power and current increase with decreasing temperature. For the \SET{} operation, the threshold voltage increases as temperature decreases, while both the threshold current and power decrease with decreasing temperature.

\section*{PCM Read-out at cryogenic temperatures}

\noindent In most phase-change materials, carrier transport is governed by thermally activated processes. At very low temperatures, however, when thermal activation is suppressed, variable-range hopping (VRH) conduction has been suggested to dominate. In addition to these characteristic mechanisms, the resistance exhibits temporal fluctuations, including resistance drift and read noise\cite{sarwat2023mechanism}. Resistance drift has a strong temperature dependence and manifests as an increase in resistance with time and temperature. On the other hand, read noise is associated with $1/f$ fluctuations\cite{MNardoneVGKarpov200904}. The temperature dependence of read noise, however, is not well documented at sub-room temperatures. In the following, we study the temperature dependence of all these characteristics. 

\subsubsection*{Charge Transport}

\noindent  The electrical resistivity of a PCM device is both temperature-dependent and strongly influenced by its phase configuration. It results from the combined contributions of the amorphous and crystalline regions, each characterized by distinct resistivities ($\rho_\text{amor}$ and $\rho_\text{cryst}$), and their respective volume fractions. These components respond differently to temperature, leading to an effective term $\rho_\text{eff}$. We find that the high-resistance states follow an Arrhenius-type behavior, with \( \rho_\text{eff}(T) \propto \exp\left(\frac{-E_\text{a,eff}}{k_\text{B}T}\right) \) where $E_\text{a,eff}$ is the effective activation energy, $k_\text{B}$ is the Boltzmann constant, and $T$ is the temperature. Notably, $E_\text{a,eff}$ increases with resistance, indicating a larger amorphous contribution to the read-out characteristics (see Supplementary Information). Furthermore, when examining the fully \RESET{} and \SET{} states as function of decreasing ambient temperature, additional observations emerge.  These are shown in Figure \ref{fig2c-IV}{} for the full \RESET{} and full \SET{}  states.  Prior to these measurements, the devices were subjected to thermal annealing to allow sufficient resistance drift due to structural relaxation to occur. We find that this treatment enables a more accurate estimation of $E_\text{a,eff}$. \\

As mentioned above, the \RESET{} state follows an Arrhenius law, meaning that the resistance value increases as the temperature is lowered. However, we observe that this behavior is valid only down to approximately \unit[180]{K}. Below this temperature, the resistance is significantly smaller than what the Arrhenius fit would predict.  Below \unit[150]{K}, the temperature dependence of the resistance can be approximately represented by the expression \( \rho_\text{eff}(T) \propto \exp\left[\left(T_0 / T\right)^{\beta}\right] \). This transition from Arrhenius-type behavior to a weaker temperature dependence suggests a departure from thermally-activated extended carriers, and is consistent with previous literature on transport in amorphous GST~\cite{MKaesMSalinga201608, DKrebsSRaoux201404, MKaesDKrebs201510}. As a result, instead of band conduction, transport becomes dominated by carrier hopping between defect states, as expected from the Mott model of VRH~\cite{NFMottEADavis2012}.  However, we note that our data does not strictly follow the Mott model prediction where $\beta = 0.25$, as can be seen from the fit attempt in the figure. This deviation can be attributed to the simplifying assumptions in the original model regarding the relationship between the maximum and most probable hopping distances. That is, the precise temperature dependence and hopping behavior can only be recovered numerically~\cite{DKrebsSRaoux201404, JMMarshallCMain200807}. Below \unit[75]{K}, we observe another transition in the transport behavior, where the resistance scaling is further arrested, i.e., the resistance values flatten or $\beta \rightarrow 0$ (see supplementary section S1). In this temperature range, it may be expected that the optimal hopping radius becomes too large due to insufficient thermal activation. This lack of temperature dependence suggests that carrier hopping must be enabled by field-driven tunneling processes.  \\

On the contrary, the \SET{}  states do not show an Arrhenius-type dependence (as shown in the inset of the same figure). Instead, the \SET{}  states exhibit a weak temperature dependence, where the resistance linearly decreases with temperature. Furthermore, this loosely follows the relationship first pointed out by Mooij~\cite{MAParkYJKim2003,CCTsuei198610} \( \rho(T) = \rho_0 + (\rho_M - \rho_0)AT \), in which $A$ is a material-dependent constant, $\rho_M$ is the resistivity corresponding to the Anderson transition, $\alpha= (\rho_M - \rho_0)A$ is the resulting temperature coefficient of resistance (TCR). It should be noted that this is an effective TCR, as it also includes the behavior of $M_\text{x}N_\text{y}$ heater. However, the heater exhibits a similarly linear but opposite (increasing resistance with temperature) characteristic (see supplementary section S2). We nonetheless observe the dependence between the extrapolated zero-kelvin resistance and the magnitude of TCR. Specifically, devices with higher starting resistance values exhibit slightly higher TCRs. This observation in real devices and in annealed films is due to Anderson localization effects induced by vacancy disorder, which has previously been observed only in macroscopic blanket GST films~\cite{HVolkerMWuttig2015, VBragagliaRCalarco201604}. \SET{} states with slightly more disordered vacancy configurations are more insulating compared to the ideal metallic conduction of the ordered \textit{hcp} phase of GST~\cite{VBragagliaRCalarco201604, JJWangWZhang201707}. \\

In summary, we find that the \RESET{} states show a strong dependence on temperature, with resistance increasing according to Arrhenius behavior and transitioning to hopping conduction at lower temperatures. In contrast, the \SET{} states exhibit a weak temperature dependence, with resistance mildly increasing as temperature decreases. These mechanisms contribute to the expanding memory window discussed in the previous section. In Supplementary Figure~S1, we also present the temperature dependence of the intermediate states. 

\subsubsection*{Resistance Drift}

\noindent In \figref{fig4-drift}{a}, we present the resistance of the device measured over time for various programmed states across the full temperature range. The measurements are carried out by applying a series of progressively increasing \RESET{} pulses. After each pulse, the device current is recorded at a constant voltage of $0.2\,\text{V}$. \figref{fig4-drift}{a} shows the normalized resistance evolution as a function of time, grouped by different temperature bins. Each resistance trace is fitted with a power-law model described by $ I(t) = I_0 \left(\frac{t}{t_0}\right)^{-\nu}$. Notably, we observe that the fitted exponent $\nu$, along with its uncertainty, enables the classification of the device's temporal behavior into three distinct, temperature-dependent regimes (see \figref{fig4-drift}{b}). This classification applies to intermediate and \RESET{} states that contain an amorphous volume, whereas in the \SET{} state the resistance-versus-time behavior is effectively temperature-independent. The first of these is the structural relaxation \textit{(SR)} regime, where amorphous phase configurations--particularly those approaching fully \RESET{} states--exhibit an increase in resistance over time, corresponding to positive values of $\nu$. For fully \RESET{} states, we observe $\nu \sim 0.12$, which is typical of GST~\cite{ABMHTalukderHSilva202402}. This regime reflects the collective, time-dependent rearrangement of atomic configurations within the amorphous volume. As a result, the activation energy for charge transport increases logarithmically with time, following \( E_\text{eff} (t) = E_\text{eff} (t_0) + \nu \cdot k_B T \ln \left(\frac{t}{t_0}\right) \). Notably, the SR regime persists across a wide temperature range in \RESET{} states, except at very low temperatures (below $130\,\text{K}$), where it is no longer evident. A second regime, referred to as partial crystallization (\textit{PC}), is characterized by negative values of $\nu$, indicating that the resistance decreases over time. This behavior is predominantly observed at high temperatures and in the intermediate resistance states. The PC regime reflects the progressive crystallization of the amorphous volume within the device. As the temperature is lowered (below \unit[350]{K}), this regime is suppressed. The third regime is observed only at very low temperatures (starting around \unit[110]{K}, and clearly at \unit[75]{K} and below), and shows the resistance increasing and decreasing following a random behavior.  We refer to this regime as random drift fluctuations \textit{(RDF)}.  Supplementary Figure~S4 shows the unbinned and unnormalized raw resistance traces, and their respective drift coefficient, for each individual temperature. \\

The \textit{PC} regime can be disregarded for cryogenic temperatures since the PCM device is expected to retain the states over extended time scales. However, at low temperatures, SR has been speculated to show an onset time \( \tau_0 \) before which no resistance drift should be observed~\cite{BKerstingASebastian2021}. The collective relaxation model~\cite{MLeGalloASebastian2018} predicts an exponential dependence of \( \tau_0 \) on \( (k_B T)^{-1} \). For example, at \( T = \unit[100]{K} \), \( \tau_0 \sim \unit[10]{sec}\), and this increases as temperature is lowered. Interestingly, as can be gathered from \figref{fig4-drift}{b-c}, this region is not evident in our measurements, even those below \unit[50]{K} (where $\tau_0 \sim \unit[1000]{years}$~\cite{BKerstingASebastian2021}). Indications of similar increases in the drift coefficient variability have also been recently observed \cite{ABMHTalukderHSilva202402}.  \\

We attribute this discrepancy to the distinction between SR and charge transport, which becomes exacerbated at low temperatures. Conventionally, SR is attributed to the removal of mid-gap defects or to bandgap widening caused by local structural reordering~\cite{BKerstingASebastian2021,DKrebsSRaoux201404,PFantiniAMani201201,SGabardiMBernasconi201508}. As long as thermally activated transport is still dominant, these effects manifest as a measurable increase in resistance. However, at very low temperatures, even if \text{SR} dynamics persist, their effects may not result in resistance drift because the dominant carrier transport mechanism transitions to VRH, and at even lower temperatures, to tunneling. The RDF regime displays pronounced resistance fluctuations in the same temperature range where tunneling transport seems to dominate. A plausible explanation is the strong sensitivity of tunneling to the spatial separation \( d \) between localized defect states, typically on the order of \unit{5-10}{ nm}. Under the WKB approximation, \( I(t) \propto \exp\left(-d(t)\right) \). Moreover, only a small number of conductive paths are expected to contribute significantly to overall device conduction. Thus, even small fluctuations in \( d \) can lead to significant changes in the tunneling probability and, consequently, device resistance. A further contributing mechanism could be electric-field induced structural relaxation triggered, which have previously been shown at higher temperatures and fields \cite{RSKhanAGokirmak202002}. \\

In summary, we observe that the resistance-time behavior exhibits a noticeable temperature dependence for phase configurations comprising an amorphous volume. Particularly, we find a separation between structural relaxation dynamics and electrical transport mechanisms at cryogenic temperatures. Under these conditions, the expected trends governed by SR are obscured because charge transport becomes dominated by direct transitions between defect states rather than by increases in activation energy. Threshold voltage onset measurements~\cite{BKerstingASebastian2021} may provide further insights into understanding these effects. 

\subsubsection*{Read Noise}

\noindent  From the same resistance-time measurements, we can also analyze the read noise characteristics as a function of both state and temperature. By subtracting the fitted resistance drift from the raw current data, we isolate the time-dependent fluctuations and treat these as read noise. From these traces, we compute the Power Spectral Density (PSD) of the current fluctuations, \(S_{II}(f)\), and examine how it varies with device state and ambient temperature. This is done by fitting the traces to the expression \( \frac{S_{II}(f)}{I^2} = \frac{Q}{f^\alpha} \) where $I$ is the average current, $Q$ is the noise amplitude parameter, and $\alpha$ is the frequency exponent. We observe that at all temperatures, and in the frequency range between \unit[0.1]{Hz} and \unit[10]{Hz}, a power-law dependence of the form $1/f^{\alpha}$ is generally present. We also extract the overall variance of the current traces $\sigma_I^2$, which enables us to compute the signal-to-noise ratio (SNR) $\frac{\sigma_I}{I}$ as a function of device resistance and temperature (see \figref{fig3-readnoise}{a}). We find that the SNR decreases with increasing resistance, in line with previous studies~\cite{SRNandakumarEEleftheriou201911}, and equivalently that it moderately decreases with decreasing ambient temperature. \\

The analysis of $\alpha$ shows that, at all temperatures, it is not strictly unity, as would be expected for ideal flicker ($1/f$) noise. Nor does it exceed 2, usually associated with Lorentzian noise generated by random-telegraph-signal fluctuations~\cite{DFugazzaALLacaita201005}. These trends are illustrated in \figref{fig3-readnoise}{b}. There is also a noticeable state dependence: the \RESET{} states tend to exhibit characteristics closer to flicker noise than the intermediate or \SET{} states. Furthermore, as the temperature is lowered, the statistical variance in $\alpha$ increases across all states, indicating greater noise heterogeneity at low temperatures. The pre-factor $Q$, which serves as a measure of variance of read fluctuations over time, also exhibits a similar temperature dependence. Notably, the spread in $Q$ increases as the temperature decreases, further confirming the trend of increasing variability at low temperatures seen in the drift coefficients. As mentioned before, at cryogenic temperatures conduction shifts into a tunneling-dominated regime. Carriers mostly flow through a small number of conductive paths, and because the tunneling probability scales exponentially with defect spatial separation, slight structural differences between programming repetitions are amplified. At higher temperatures, many parallel, thermally activated paths average those differences out, so $Q$ collapses into a narrower distribution.\\

In summary, we find that the noise behavior still exhibits characteristics consistent with flicker-type noise at low ambient temperatures. We observe no significant temperature dependence in the PSD slope or in the nature of the transitions that produce this behavior. The PSD amplitude pre-factor $Q$ becomes significantly more variable at cryogenic temperatures, with higher variability for all intermediate and \RESET{} states.

\section*{Compute performance}

\noindent Based on the temperature-dependent measurements described above, we derive empirical models that relate various device metrics to device resistance and ambient temperature. These models enable us to explore the collective behavior of PCM devices, particularly in the context of IMC. For example, in tasks such as matrix-vector multiplication (MVM), expressed as $y=Wx$, where $x$ is the input vector and $W$ is the weight matrix encoded in device conductance. For this, we developed a hardware emulator that captures key device behaviors and intrinsic error sources, while excluding peripheral circuitry\cite{aihwkit_nonidealities}. Using this emulator, we analyze the MVM accuracy over a one-year timescale (see implementation in supplementary S4). For a $256\times256$ matrix with inputs randomly sampled from a normal distribution this is illustrated in \figref{fig5-sim}{a and b}.\\

Panel \textit{a} of the figure illustrates the temporal evolution of stored weights, which we find to be strongly dependent on the ambient temperature. At room temperature (\unit[300]{K}), resistance drift dominates, causing a gradual and proportional decrease in weight values over time. At intermediate temperatures (\unit[100--200]{K}), both resistance drift and temporal fluctuations contribute significantly, resulting in a broader spread in the weight distribution. At cryogenic temperatures approaching \unit[5]{K}, resistance drift is largely arrested; however, the weights become increasingly susceptible to random fluctuations. Panel~\textit{b} quantifies the impact of these effects on compute accuracy, expressed as $E_{\ell_2}(t) = \frac{\Vert \tilde{y}(t) - y \Vert_2}{\Vert y \Vert_2}$. Accuracy degrades with decreasing temperature in the intermediate range, with compute errors at \unit[200]{K} and \unit[100]{K} being larger than that at room temperature. Interestingly, at lower temperatures near \unit[5]{K}, accuracy comparable to room temperature is recovered. This suggests that while increased read noise penalizes compute performance, it is partially compensated by the suppression of drift and reduced state dependence. \\

The large memory window available at low temperatures enables weight programming within a restricted conductance range by clipping the maximum conductance to a value $G_\text{max}$ lower than the full-\textit{SET} conductance.  This is because the minimum weight programmable in the crossbar (after appropriate rescaling) is $|w_\text{min}| \propto G_\text{min} / G_\text{max}$.
This reduction in $G_\text{max}$ lowers the average column current and the corresponding IR-drop due to line resistance, and reduces power dissipation in the crossbar. These effects are expected to be beneficial for the scalability of larger arrays. The results are illustrated in \figref{fig5-sim}{c}. Notably, at \unit[5]{K}, a four-decade reduction in $G_\text{max}$ increases the one-year error by only a factor of two. This allows operation with significantly reduced column current---and hence lower IR-drop---while still maintaining acceptable compute accuracy. However, this trade-off becomes less favorable at higher temperatures: the same reduction in $G_\text{max}$ leads to a much more pronounced degradation in accuracy. At room temperature, the usable conductance window narrows, limiting effective clipping and thereby reducing computational precision.  It is also important to recognize that the weight-conductance mapping is fundamentally limited by the read noise $\sigma_{\text{read}}$. While the memory window may expand at cryogenic temperatures, increased read noise in such conditions introduces a competing constraint. Thus, the number of distinguishable conductance levels is ultimately governed by the interplay between dynamic range and read noise.\\

Taken together, the results indicate that for in-memory computing applications, cryogenic PCM devices offer their most significant advantages at low temperatures, where the combination of a large resistance window and effectively arrested drift enhances performance. In the upper end of the cryogenic temperature range, although the resistance window remains wide, compute accuracy becomes increasingly limited by structural relaxation. Further insights will be necessary to fully position the performance metrics across temperature regimes. For instance, the use of global drift compensation and iterative programming schemes, supported by peripheral circuits that can accommodate large resistance variations can help reduce accuracy degradation~\cite{VJoshiEEleftheriou202005,MJRaschVNarayanan202308}. These results, nonetheless, highlight an important finding for practical applications. They show that while certain performance metrics improve at lower temperatures, others degrade. The interplay between them ultimately governs the IMC behavior under cryogenic operation.

\section*{Conclusions and outlook}

\noindent In summary, we have systematically characterized the electrical behavior of prototypical PCM devices down to \unit[5]{K} for in-memory computing applications. Our measurements provide evidence that the fundamental electro-thermal mechanisms underlying \RESET{} and \SET{} programming remain valid even at these ultra-low temperatures, though with notable quantitative shifts attributable to changes in material properties under cryogenic conditions. Our analysis of read-out characteristics reveals transitions in the dominant transport mechanisms as temperature decreases--from band conduction at higher temperatures to hopping conduction below \unit[180]{K}, and eventually to tunneling-dominated transport below \unit[75]{K}. Resistance drift measurements further indicate a shift in the nature of temporal fluctuations across temperatures. Using empirical models fitted to our data, we demonstrate that compute accuracy at cryogenic temperatures is governed by the interplay of temperature-dependent changes across multiple device characteristics. These results provide important fundamental insights into a temperature regime that has growing relevance for  PCM-based in-memory computing.

\section*{Methods}

\small

\noindent \textit{Device fabrication}. The PCM devices studied were mushroom-type cells fabricated using standard thin-film deposition techniques. The device stack comprised an \unit[80]{nm} thin film of \GST{} as the phase-change material, sandwiched between a bottom metal-nitride ($M_xN_y$) electrode and a top metal electrode, with the bottom electrode acting as a heater. All materials were sputter-deposited. The bottom electrode diameter was approximately \unit[36]{nm}, defining the active area. The lateral device dimensions were nominally \unit[200]{nm} by \unit[540]{nm}. \\

\noindent \textit{Cryogenic measurement setup}. Electrical characterization at cryogenic temperatures was performed using two separate cryogenic setups, one cooled by liquid nitrogen (LN2) and the other by liquid helium (LHe). Both setups utilized a Janis ST-500-2-UHT cryogenic probe station integrated with a Lakeshore 336 temperature controller. Electrical measurements employed a Keithley 2636B Source Measure Unit (SMU) for high-accuracy current-voltage characterization, an Agilent 81150A Arbitrary Waveform Generator (AWG) for programming pulses, and a Tektronix DPO4104 oscilloscope for pulse characterization. Device contacts were made using Cascade Microtech Dual-Z Ground-Signal-Signal-Ground probes. \\

\noindent \textit{Measurement procedures}. Prior to electrical characterization, the devices underwent a standardized wake-up procedure to ensure reproducible performance and minimize effects related to material segregation or device fatigue. Each device was subjected to bursts of $10^5$--$10^6$ \RESET{} pulses at an intermediate amplitude (\unit[2]{V}). During this conditioning step, programming curves were recorded at exponentially increasing intervals to monitor potential fatigue-related resistance variations. A stable operating regime was reached, indicated by saturation of the \SET{} and \RESET{} resistances and consistent \RESET{} currents across successive measurements. Programming characteristics were obtained by applying incrementally increasing \RESET{} voltage pulses while ensuring a consistent initial \SET{} state between pulses. Pulse amplitudes typically ranged from \unit[1--2.5]{V}, with pulse durations of \unit[50]{ns} with \unit[8]{ns} rise/fall time for \RESET{} operations, and approximately \unit[1]{$\mu$s} rise/fall for \SET{} pulses. Device voltages and currents were digitized simultaneously using the oscilloscope with high-impedance and low-impedance configurations, respectively. From these waveforms, the programming power threshold required to initiate melting (\RESET{}) was extracted by identifying the point where device resistance reached twice its baseline \SET{} resistance at a given temperature. Static read-out characteristics were obtained through low-bias current-voltage (IV) measurements performed between \unit[300]{K} and \unit[5]{K}. Multiple devices were initially programmed at room temperature (\unit[300]{K}) in different states (SET, intermediate, RESET), after which they underwent an annealing step at \unit[315]{K} overnight and then relaxed at room temperature for 30 days to stabilize the resistance states. For the IV characterization, at each measurement temperature, voltage sweeps up to $\pm\unit[0.4]{V}$ were conducted using the SMU while being careful to prevent unintentional switching. Resistance drift measurements were conducted by applying \RESET{} pulses and immediately monitoring current over extended time periods (at least three orders of magnitude in time from the moment of first read). Device currents at a constant bias voltage (\unit[0.2]{V}) were recorded, and resistance-time data were fitted using power-law expressions, capturing structural relaxation dynamics and other temporal fluctuations. Special care was taken to minimize the initial measurement delay. Noise characteristics were extracted by subtracting fitted drift from raw data, yielding residual fluctuations analyzed through PSD fitting. The slightly higher-than-expected slope observed in the \RESET{} state ($\alpha \approx 1.2$ instead of $\alpha \approx 1$) may be caused by additional low-frequency environmental noise being picked up by the cryostat itself.

\section*{Acknowledgments}

\noindent This work is supported by the IBM Research AI Hardware Center. We would
like to thank Timothy Philicelli, Valeria Bragaglia, and Vara Prasad Jonnalagadda for technical discussions. We thank Cezar Zota for access to the liquid helium cryogenic probe station.

\section*{Competing financial interests}
\noindent The authors declare no competing financial interests.

\section*{Data availability}
\noindent The data that support the findings of this study are available from the corresponding author upon reasonable request.

\section*{Contributions}
\noindent D.G.F.L. carried out the electrical characterization of devices and data analysis. S.G. contributed to the measurements scripts. A.F. assisted with the measurement setup. M.L. provided input on the simulation studies. G.S.S. and A.S. defined the research question. G.S.S. supervised the project. D.G.F.L. and G.S.S. wrote the manuscript with input from all authors.

\section*{References}
\bibliographystyle{naturemag}
\bibliography{References}

\newpage
\section*{Figures}

\begin{figure*}[h]
   \includegraphics{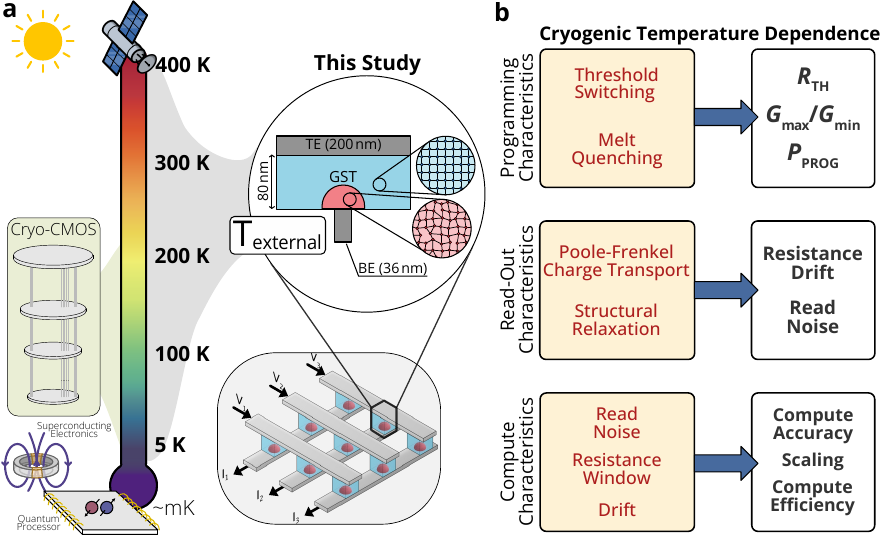}
    \centering
    \caption{a) An illustration showing that phase change memory-based in-memory computing can be applied across a wide temperature span. b) In this study, we characterize the temperature dependent behavior of mushroom-type phase-change memory devices in this temperature range (from \unit[400]{K} down to \unit[5]{K}).  Specifically, we systematically investigate the programming and read-out behaviors, analyzing and reasoning their temperature-dependencies.}
    \label{fig1-intro}
\end{figure*}

\begin{figure*}
    \includegraphics{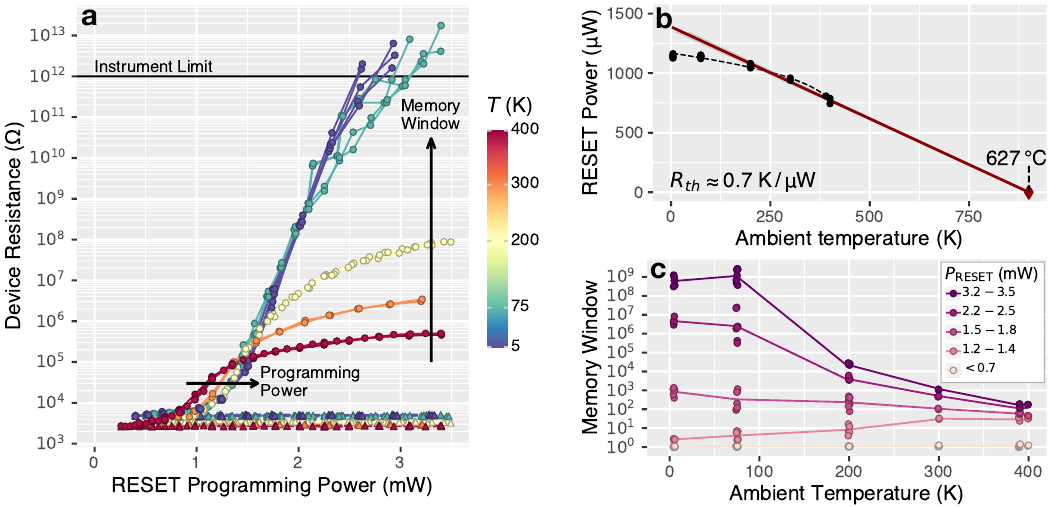}
    \centering
    \caption{a) RESET resistance as a function of programming energy measured at ambient temperatures of \unit[5]{K}, \unit[75]{K}, \unit[200]{K}, \unit[300]{K}, and \unit[400]{K}. Triangular points show the starting SET state for each pulse. b) RESET onset power as a function of ambient temperature. c) Memory Window, defined as $R_\text{RESET} / R_\text{SET}$ for various RESET pulse amplitudes as a function of ambient temperature.}
    \label{fig2-prog}
\end{figure*}

\begin{figure*}
    \includegraphics{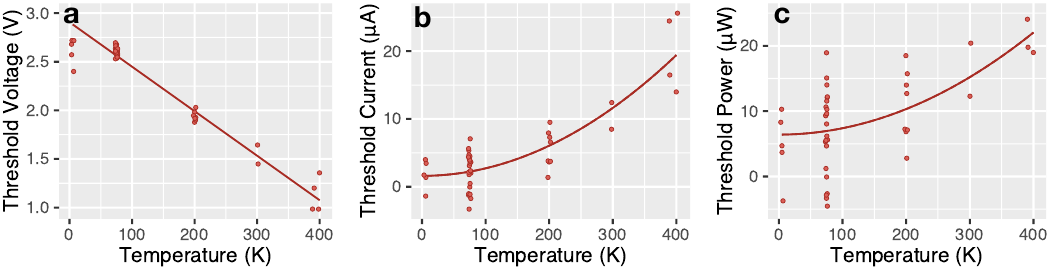}
    \centering
    \caption{Temperature dependence of threshold switching parameters of the full RESET state: threshold voltage (a), threshold current (b), and threshold power (c). }
    \label{fig2b-thresh}
\end{figure*}

\begin{figure*}
    \includegraphics[]{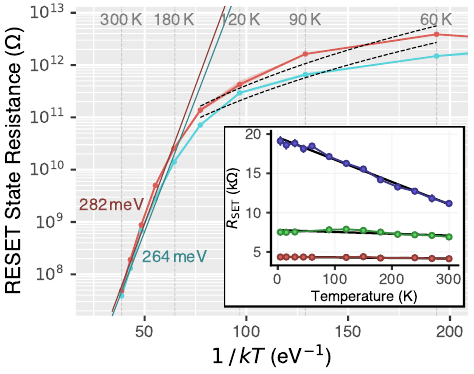}
    \centering
    \caption{RESET state resistance at read voltage of \unit[0.2]{V} of two devices programmed at room temperature and progressively cooled.  Activation energies of the Arrhenius law are highlighted. Fit with $T^{-0.25}$ VRH law between \unit[150]{K} and \unit[60]{K} is shown as a black dashed line. Inset: SET state resistance at read voltage of \unit[0.2]{V} of three devices programmed at room temperature and progressively cooled.}
    \label{fig2c-IV}
\end{figure*}

\begin{figure*}
   \includegraphics{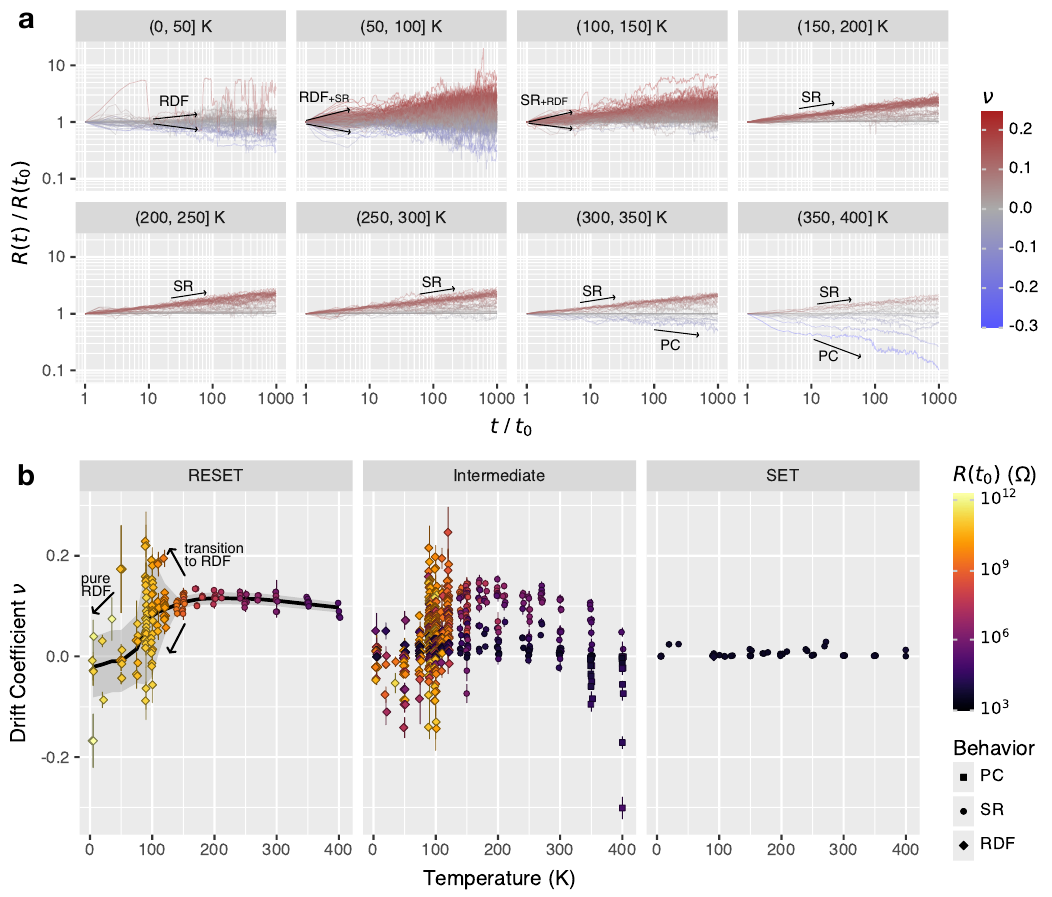}
    \centering
    \caption{a) Normalized resistance-time plot of many traces across all device states. The color of the trace corresponds to the value of the drift coefficient fitted to it. The horizontal axis is in units of time since the first read $t_0$, and the vertical axis shows the resistance normalized to the value during the first read, $R(t_0)$.
    b) Drift coefficient extracted from the traces as a function of temperature, separated into RESET, intermediate, and SET states.
    Different regimes as a function of temperature are highlighted as Partial Crystallization (PC), Structural Relaxation (SR), and Random Fluctations (RDF).} 
    \label{fig4-drift}
\end{figure*}

\begin{figure*}
   \includegraphics{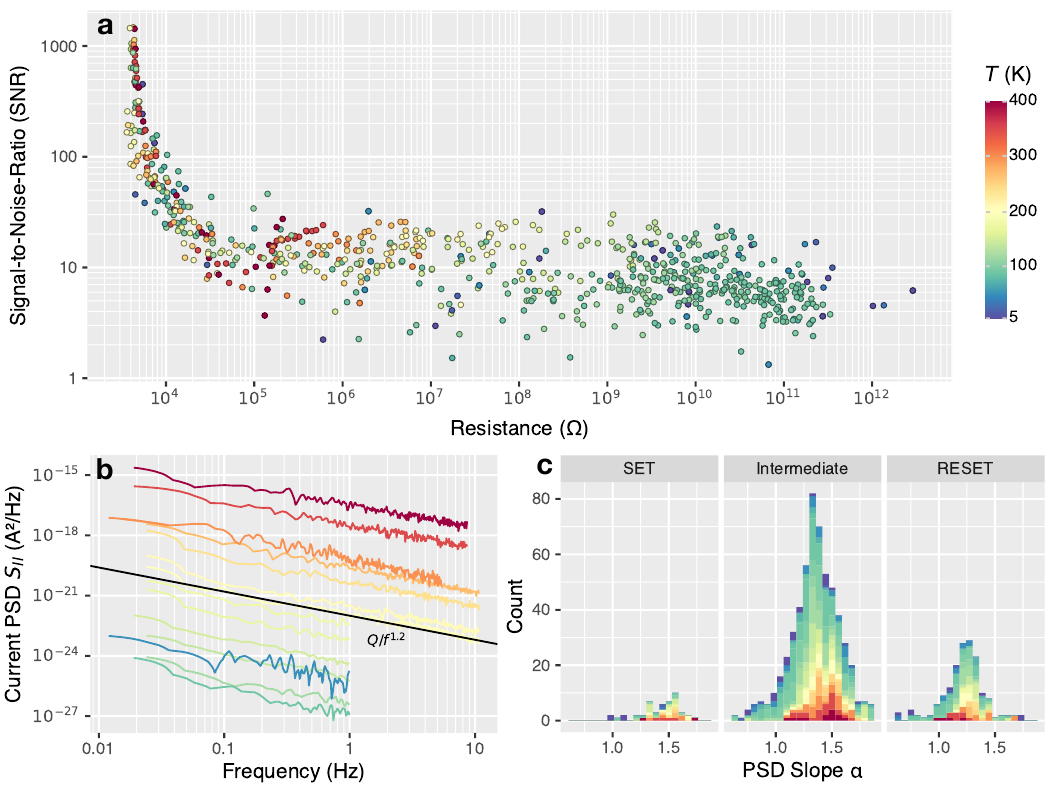}
    \centering
    \caption{a) Signal-to-Noise-Ratio as a function of device resistance. b) Small selection of current PSDs in the RESET state for various temperatures, showing a median slope $\alpha \approx 1.2$, and (c) statistics for PSD slope. Slopes are fitted for $f < \unit[1]{Hz}$.}
    \label{fig3-readnoise}
\end{figure*}

\begin{figure*}
   \includegraphics{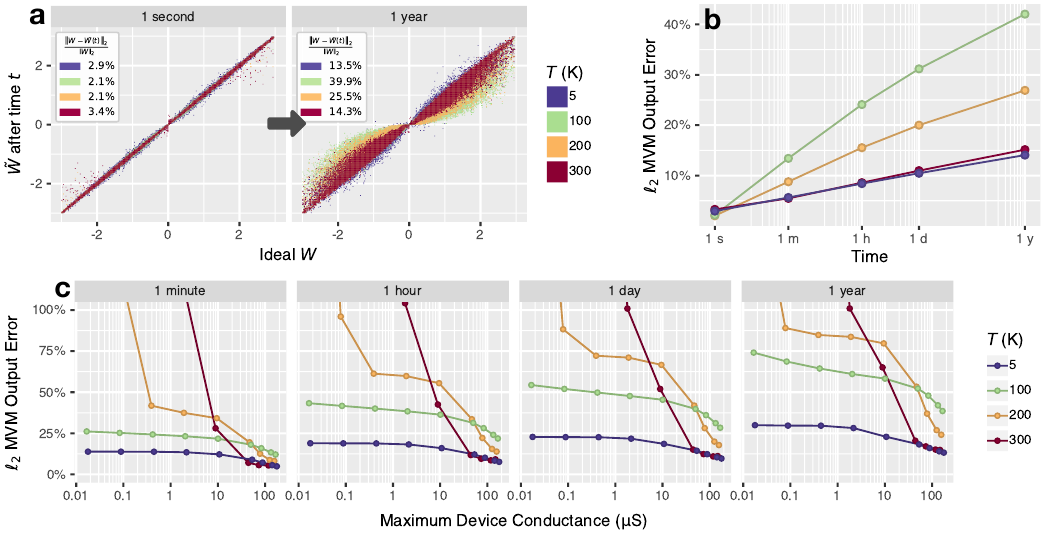}
    \centering
    \caption{Simulation of MVM compute accuracy as a function of temperature, computed for \unit[300]{K}, \unit[200]{K}, \unit[100]{K}, and \unit[5]{K}. a) Evolution of the weight matrix stored in the simulated crossbar's devices' conductance from 1 second to 1 year after programming. Legend shows 2-norm distance between target and stored matrices, normalized to the 2-norm of the target matrix. b) Evolution of average output error from ideal target computed from randomly-sampled inputs. c) Evolution of average output error from 1 minute to 1 year after programming, as a function of maximum programmed $G_\text{max}$ in the crossbar.}
    \label{fig5-sim}
\end{figure*}

\end{document}


\title{Cryogenic In-Memory Computing with Phase-Change Memory}

\author{Davide G. F. Lombardo}
\affiliation{IBM Research -- Europe, S\"{a}umerstrasse 4, 8803 R\"{u}schlikon, Switzerland}
\affiliation{École polytechnique fédérale de Lausanne (EPFL), Lausanne, Switzerland}
\author{Siddharth Gautam}
\affiliation{IBM Research -- Europe, S\"{a}umerstrasse 4, 8803 R\"{u}schlikon, Switzerland}
\affiliation{École polytechnique fédérale de Lausanne (EPFL), Lausanne, Switzerland}
\author{Alberto Ferraris}
\affiliation{IBM Research -- Europe, S\"{a}umerstrasse 4, 8803 R\"{u}schlikon, Switzerland}
\affiliation{École polytechnique fédérale de Lausanne (EPFL), Lausanne, Switzerland}
\author{Manuel Le Gallo}
\affiliation{IBM Research -- Europe, S\"{a}umerstrasse 4, 8803 R\"{u}schlikon, Switzerland}
\author{Abu Sebastian}
\affiliation{IBM Research -- Europe, S\"{a}umerstrasse 4, 8803 R\"{u}schlikon, Switzerland}
\author{Ghazi Sarwat Syed}
\affiliation{IBM Research -- Europe, S\"{a}umerstrasse 4, 8803 R\"{u}schlikon, Switzerland}
\email{ghs@zurich.ibm.com, ase@zurich.ibm.com}

\maketitle

\section{More Insights into the Current-Voltage Characteristics}

\noindent In section 2 of the main text, we have discussed the readout properties of the SET and RESET states. Here we expand on these results. \\

\noindent To measure the $I(V, T)$ characteristics of the devices as a function of the phase configuration, 16 devices in total were programmed at room temperature: 4 full SET devices, 4 full RESET devices, and 8 intermediate states. To ensure that the vast majority of drift would happen before the devices would enter the cryostat, they were first heated to $315\,\text{K}$ overnight, and then were left under room conditions to further relax for about a month. \\

\begin{figure}[h]
    \centering
    \includegraphics[width=0.5\linewidth]{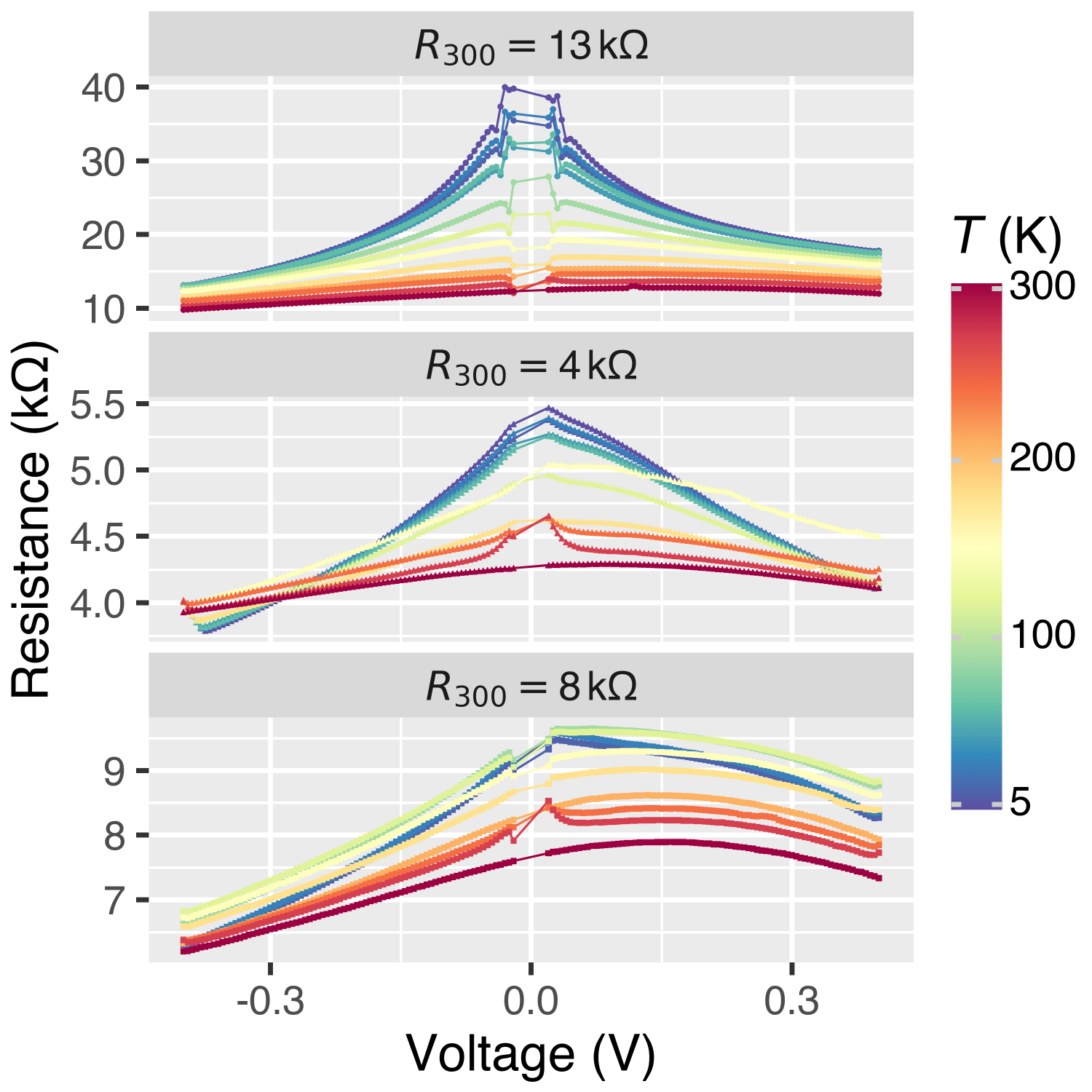}
    \caption{R-V sweeps of the three devices in the SET state shown in the main text.}
    \label{fig:IV-set}
\end{figure}

\begin{figure}[h]
    \centering
    \includegraphics[height=0.33\linewidth]{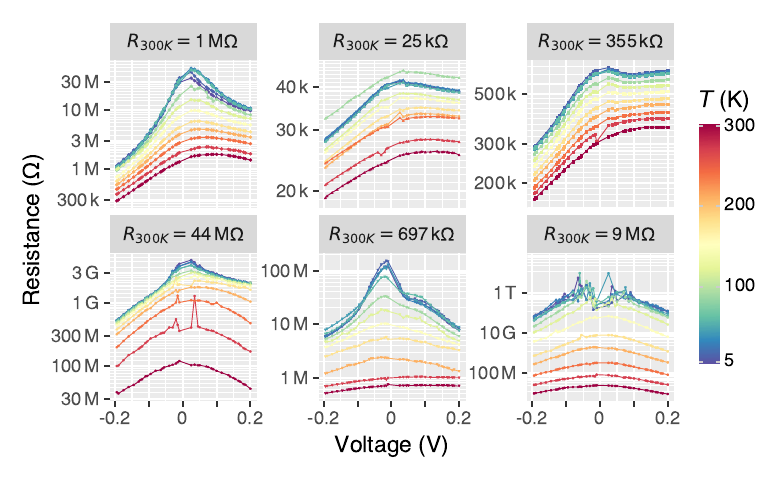}
    \includegraphics[height=0.33\linewidth]{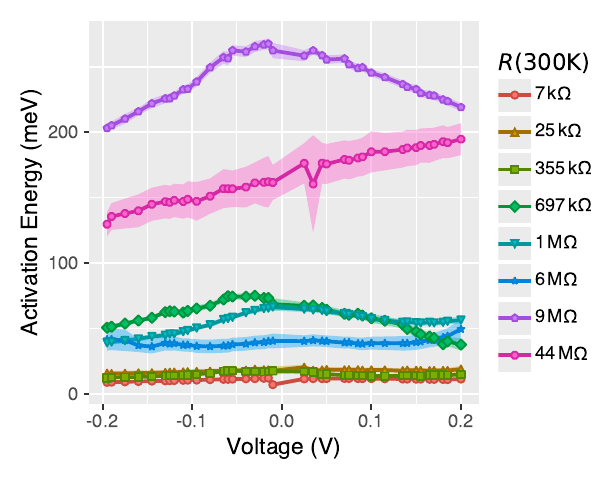}
    \caption{(left) R-V sweeps of 6 devices programmed into various intermediate states. (right) Effective activation energy of the devices as a function of read voltage.}
    \label{fig:IV-interm}
\end{figure}

\begin{figure}[h]
    \centering
    \includegraphics[height=0.25\linewidth]{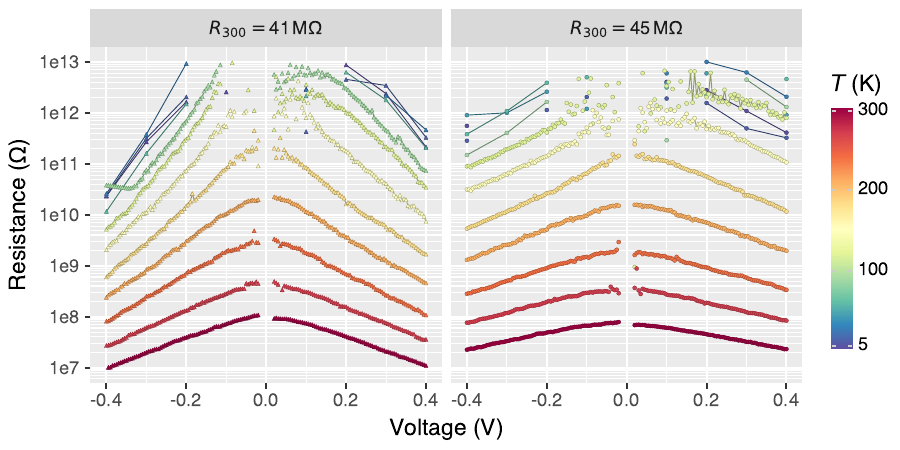}
    \includegraphics[height=0.25\linewidth]{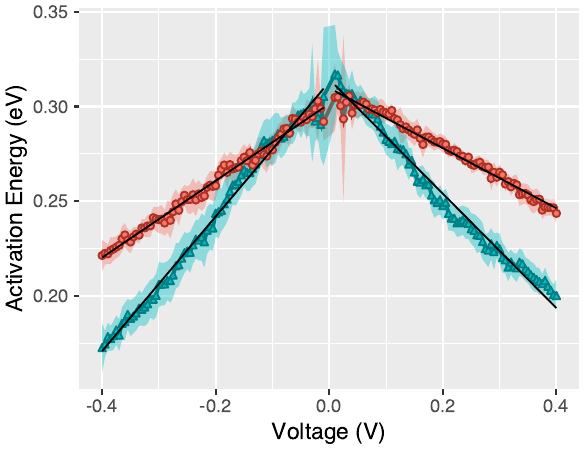}
    \includegraphics[height=0.3\linewidth]{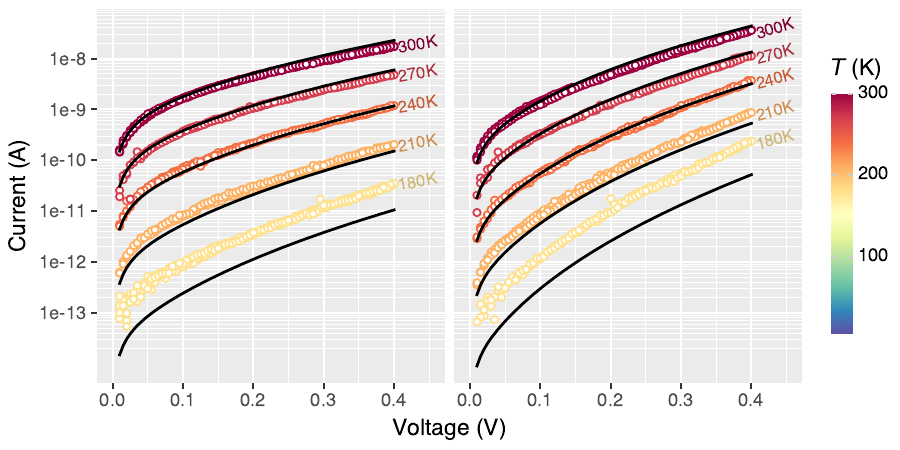}
    \includegraphics[height=0.3\linewidth]{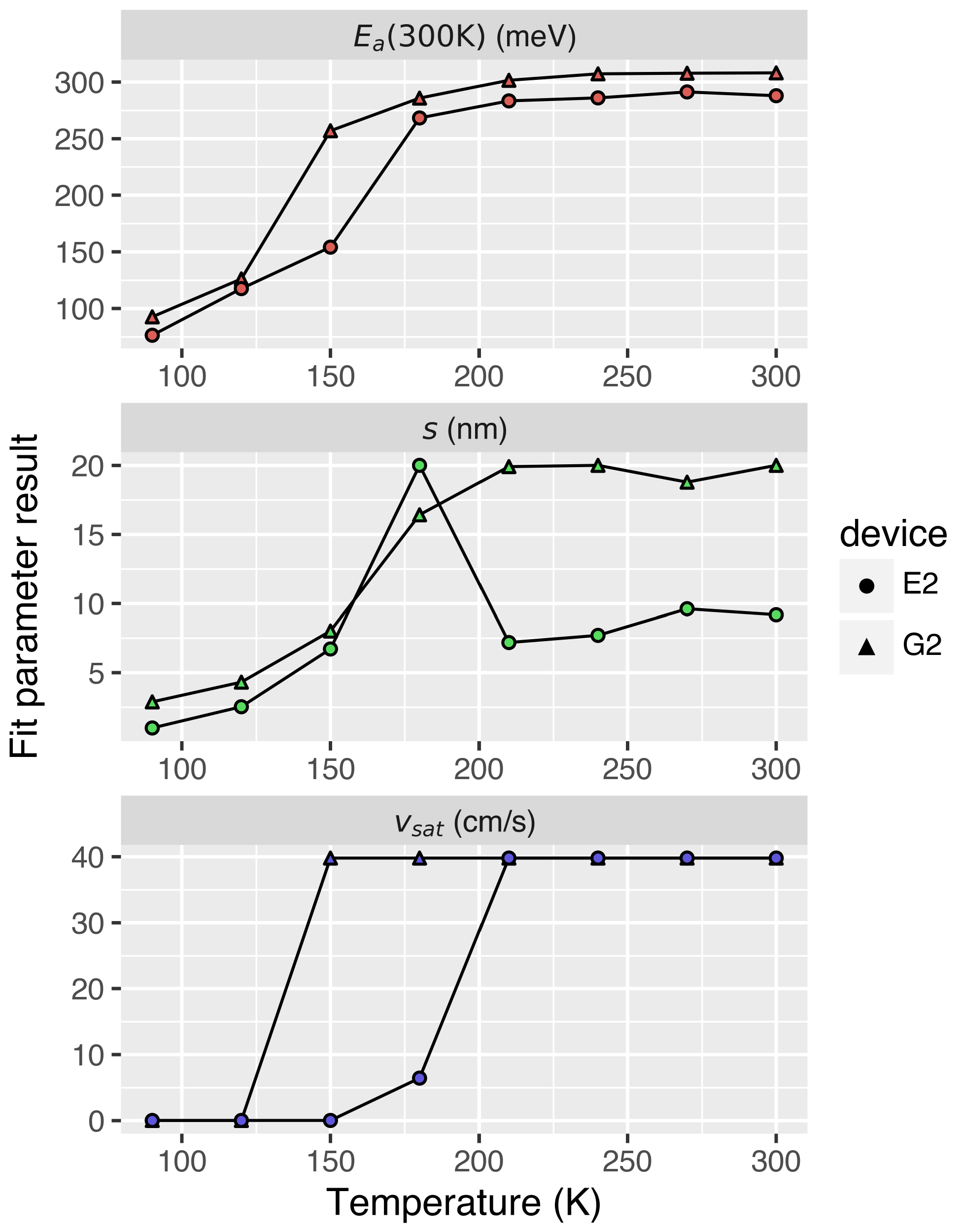}
    \caption{Top left: R-V sweeps at various temperatures for two devices in full RESET state. Top right: Effective activation energy $E_{a,\text{eff}}$ of the devices as a function of read voltage. Bottom left: 3D Poole-Frenkel fit of the I-V characteristics for the two devices. Bottom right: Activation energy $E_{a}$, inter-trap separation distance $s$ and carrier saturation velocity $v_\text{sat}$ extracted from the 3D Poole-Frenkel I-V fit as a function of temperature.}
    \label{fig:IV-reset}
\end{figure}

All Resistance-Voltage (R-V) sweeps acquired for devices in the SET, intermediate and RESET states are shown, respectively, in Figure~\ref{fig:IV-set}, Figure~\ref{fig:IV-interm}, and Figure~\ref{fig:IV-reset}.
Additional analysis for the intermediate and RESET states is also shown. We extracted the effective activation energy at fixed read voltages. We note a rough correlation between room temperature resistance and activation energy, in addition to a wide display of field sensitivity. These are a consequence of the complex dependency of electrical transport on the amorphous phase configuration of the device in the intermediate states. \\

In Figure~\ref{fig:IV-reset} we also show our model for the high-temperature charge transport characteristics in two RESET state devices with a 3D Poole-Frenkel model \cite{MLGalloDKrebs201509}. We see that the model is able to reproduce the devices' I-V characteristics with physically reasonable parameters, but only above approximately \qty{200}{\kelvin}.
At \qty{180}{\kelvin} and below, the current predicted by the model falls short of the experimental data. Simultaneously, parameters like activation energy, average inter-trap distance and carrier saturation velocity start deviating significantly from their room-temperature values. This points to $\sim200 \pm \qty{10}{\kelvin}$ as the temperature at which localized carrier transport mechanisms start coming into play. \\

\begin{figure}[h]
    \centering
    \includegraphics[width=1\linewidth]{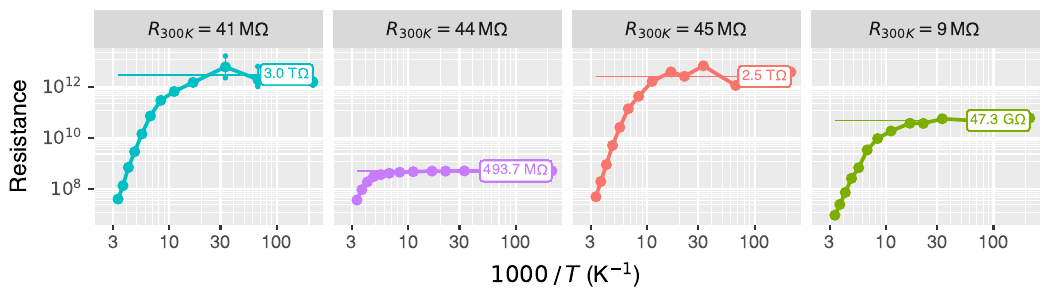}
    \caption{Full Arrhenius plot of 4 different devices programmed in the RESET state from \SI{300}{K} to \SI{5}{K}. Label and line show the average saturation resistance reached below \SI{75}{K}.}
    \label{fig:arrhenius-reset}
\end{figure}

We also note that resistance at \qty{75}{\kelvin} and below seems to saturate. This can be seen in Figure~\ref{fig:arrhenius-reset}, which shows the resistance vs inverse temperature of four devices programmed to their respective maximum RESET state at room temperature. They all saturate at some point between \qty{100}{\kelvin} and \qty{50}{\kelvin}, and the saturation value does not seem to be obviously correlated with the room temperature resistance.
This again shows low temperature transport to be tunneling based rather than temperature-activated, and furthermore that it strongly depends on the amorphous phase configuration.

\clearpage

\section{Insights into the Temperature dependency of the bottom electrode}

\noindent In this section we discuss the effect of the bottom electrode (heater) on the programming, and the readout behavior of devices shown in the main text. \\

The electrical and thermal conductivities of the $M_\text{x}N_\text{y}$ heater are temperature dependent, and its room temperature electrical resistance is measured to be comparable to the resistance of the device in the melt state. A considerable fraction of electrical power is therefore being dissipated in the heater itself, lowering the overall thermal resistance $R_\text{TH}$ of the device. This complicates both interpretation and quantitative modeling of the temperature dependence of programming power and SET resistance. Correcting for the heater's effect would require estimating the temperature coefficient of resistivity of the electrode material. To estimate it, we have measured the I-V characteristics of two devices with a geometry which allowed measuring the $M_\text{x}N_\text{y}$ heater material in isolation. It behaves as an ideal ohmic resistor, with flat response over a \qty{-0.8}{\volt} to \qty{+0.8}{\volt} sweep, at all temperatures examined. \\

\begin{figure}[h]
    \centering
    \includegraphics[width=0.5\linewidth]{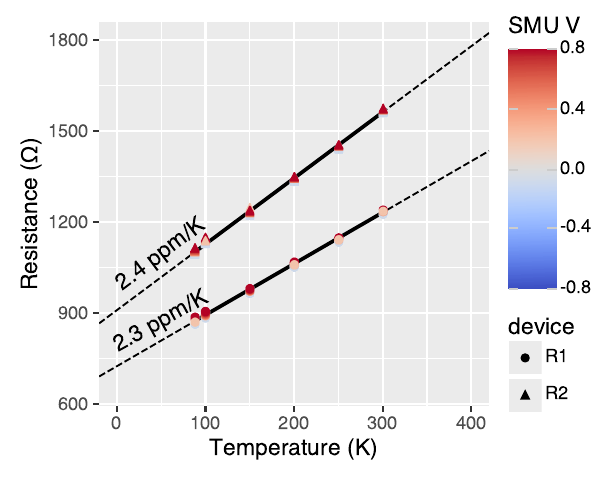}
    \caption{Resistance vs temperature of two modified electrode-only devices.}
    \label{fig:heater-electrode}
\end{figure}

The data is shown in Figure~\ref{fig:heater-electrode}. While it is difficult to apply these results directly due to the geometrical difference between these modified devices and the ones under study, this shows that the heater electrode has the expected metallic temperature dependence.

\clearpage

\section{More Insights into Resistance Drift Behavior}\label{subsec:drift-suppl}
\noindent This section presents the full drift dataset shown in the main text.\\ 

For almost all traces, the time elapsed between the programming pulse and the first resistance read $t_0$ is 2 power line cycles, i.e. \qty{40}{\milli\second}.
For some low-temperature resistance traces ($T < \qty{75}{\kelvin}$), the first read time was increased up to $t_0 = \qty{1}{\second}$ to get more accurate readings. No qualitative or quantitative difference is observed between these traces and the rest of the dataset, and they are therefore bunched together in this figure.
Regardless of the value of $t_0$, the resistance was monitored for long enough so that all traces cover at least 3 orders of magnitude in time.
\begin{figure}[h]
    \centering
    \includegraphics[width=0.95\linewidth]{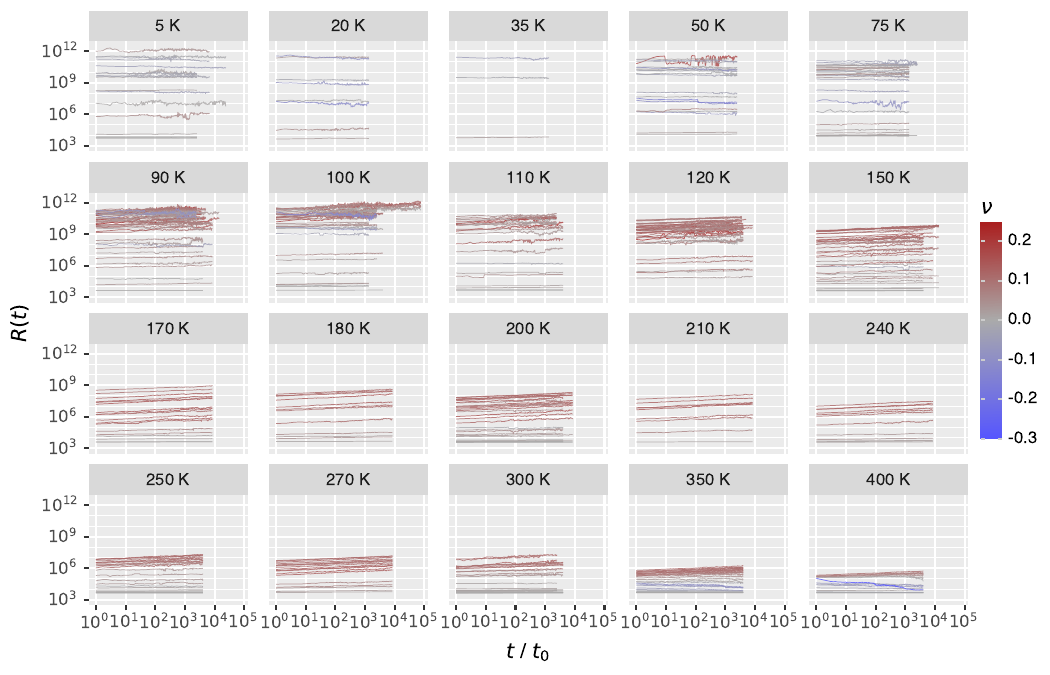}
    \includegraphics[width=0.95\linewidth]{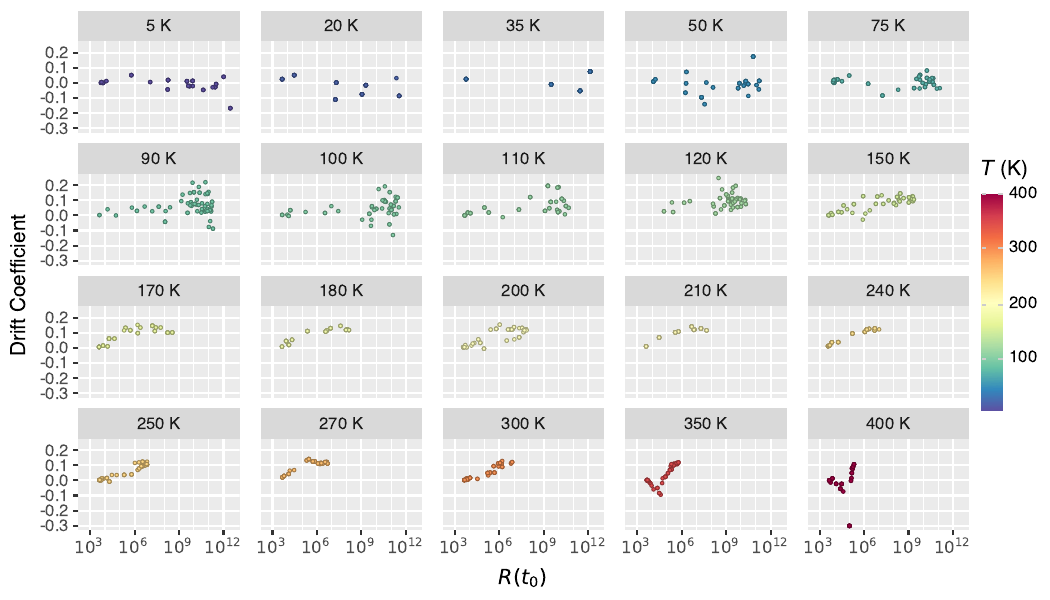}
    \caption{Top: Resistance over time, with time normalized to units of time between programming pulse and first read. Different subpanels show corresponding temperatures. The color of the trace corresponds to its fitted drift coefficient $\nu$.
    Bottom: Drift coefficient versus resistance at first read extracted from the traces in the top panel. Each subpanel shows a different temperature.}
    \label{fig:drift-suppl}
\end{figure}

\clearpage

\section{More Insights into the Simulator Architecture }\label{subsec:sim-architecture}

\noindent In section 3 of the main text, we have presented our results for the temperature dependence of the devices' analog compute performance. In this section, we describe in detail the architecture and implementation of the simulator software, including the modeled device non-idealities. \\

\noindent The MVM simulator used in the study is a simple matrix-vector multiplier
where each matrix element is mapped to the conductance values of corresponding PCM devices.
The input signals, encoded into voltage amplitudes,
and the summation of the output signals for the final result
are achieved in a straightforward manner through the crossbar architecture.
This simple electrical scheme maps to the following operation:
\begin{align}
  y_i &= \sum_j M_{ij} x_j \;\to\; \tilde{y}_i = \sum_j \tilde{M}_{ij} x_j ,\notag\\[0.3em]
  I_i &= \sum_j G_{ij} V_j \;\to\; \tilde{I}_i = \sum_j \tilde{G}_{ij} V_j .
  \label{eq:crossbar}
\end{align}
where \(M \in \mathbb{R}^{n \times n}\), with \(n = 256\), and
\begin{equation}
  \tilde{G}_{ij} = G_{ij}\bigl(T, V_j, t\bigr) + \sigma_r\!\bigl(G_{ij}\bigr),
  \label{eq:nonidealities}
\end{equation}
takes into account ambient temperature, bias voltage, drift, and read noise.
 The \(x \mapsto V\) and \(I \mapsto y\) conversions are abstracted into the following scalar maps:
\begin{align}
  V(x_j) &= \alpha \cdot x_j \hspace{5em} & y(I_i) &= \omega \cdot I_i ,\notag\\
  \alpha &= \frac{V_{\max}}{x_{\max}}        & \omega &= \frac{M_{\max} x_{\max}}{G_{\max} V_{\max}}
\end{align}
These give the desired result \(\vec{y} = M \vec{x}\) when
\begin{equation}
  G_{ij} = \frac{G_{\max}}{M_{\max}}\, M_{ij}
\end{equation}
In a real chip, the \(x \mapsto V\) and \(I \mapsto y\) maps are implemented by row-wise and column-wise multipliers in the periphery of the crossbar. The ``negative conductance'' required for all \(M_{ij} < 0\) is similarly implemented through peripheral circuitry. Since operations such as input rescaling, digital-to-analog conversion, analog-to-digital conversion, subtraction, and output rescaling operations are all performed by the digital peripheral circuitry, they are not considered sources of error in the simulator. The simulator can be summarized by the pseudo-code in Algorithm~\ref{alg:simulator}. More insights into such a framework is documented in Ref \cite{aihwkit_nonidealities}. 

\begin{algorithm}[ht]
\caption{Simulator outline}\label{alg:simulator}
\SetAlgoLined
\DontPrintSemicolon
generate weight matrix \(M\in\mathbb{R}^{256\times256}\) by sampling \(\mathcal{N}(M_{ij})\) for all \(i,j\)\;
generate input vector \(x\in\mathbb{R}^{256}\) by sampling \(\mathcal{N}(x_j)\) for all \(j\)\;
compute target \(y \leftarrow M x\)\;
map \(x\) to voltage vector \(V \leftarrow \alpha x\)\;
map \(M\) to device matrix \(G \leftarrow \dfrac{G_{\max}}{M_{\max}}\,M\)\;
\For{each non-ideality}{
  compute \(\tilde{G} \leftarrow G\big|_{\text{non-ideality}(T,t)}\)\;
  compute \(\tilde{I} \leftarrow \tilde{G} V\)\;
  compute \(\tilde{y} \leftarrow \omega \tilde{I}\)\;
}
compare \(y\) and \(\tilde{y}\left(T,t\right)\)\;
\end{algorithm}
\noindent The bulk of the simulator's code is the implementation of the conductance non-idealities, i.e., implementing Eq. \eqref{eq:nonidealities}.
Table~\ref{tab:metrics} shows the numerical inputs and outputs from the simulator which are visualized in the main text.

\subsection*{Finite Conductance Window}
\begin{figure}[h]
    \centering
    \includegraphics[width=0.45\linewidth]{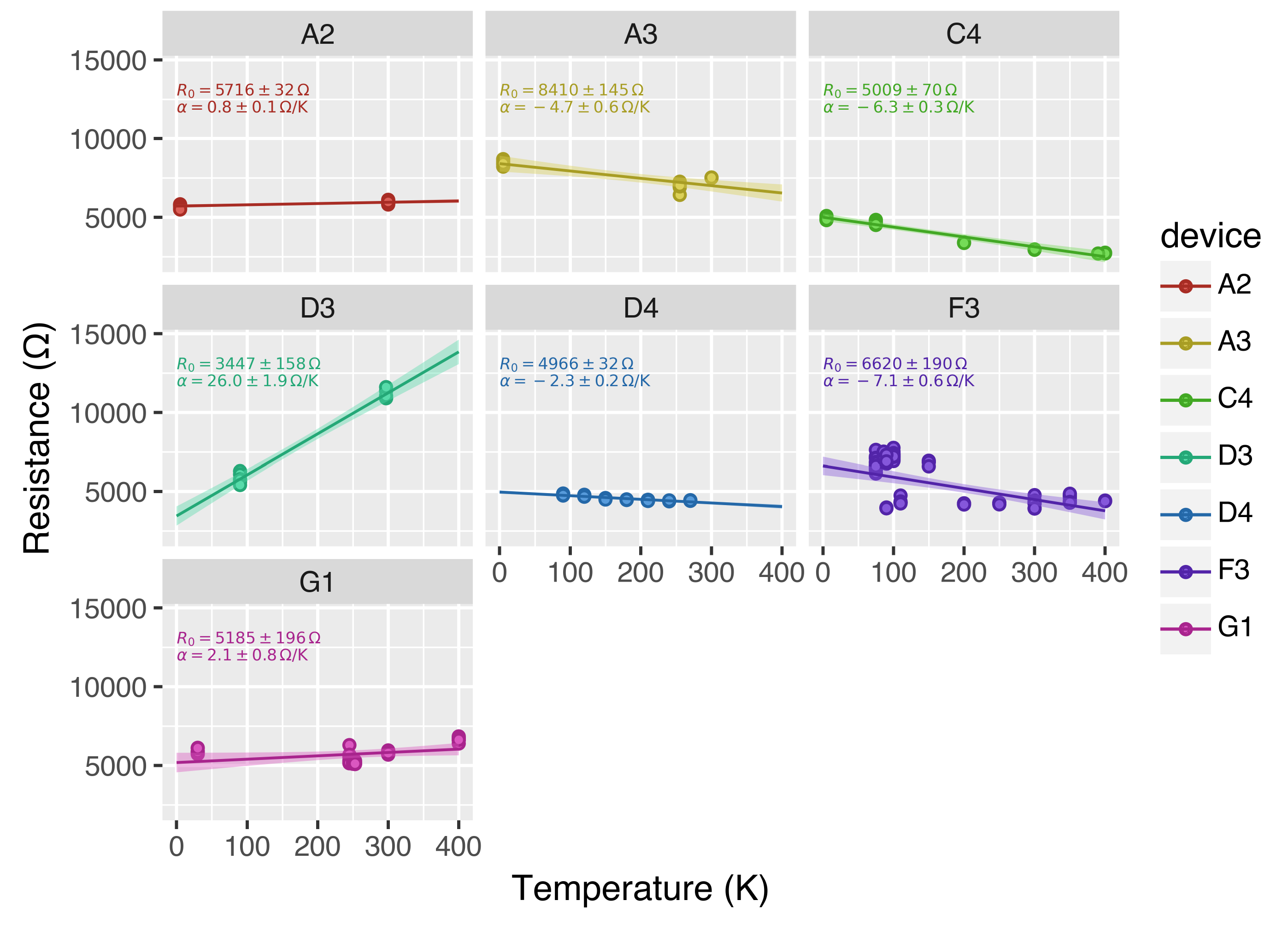}
    \includegraphics[width=0.45\linewidth]{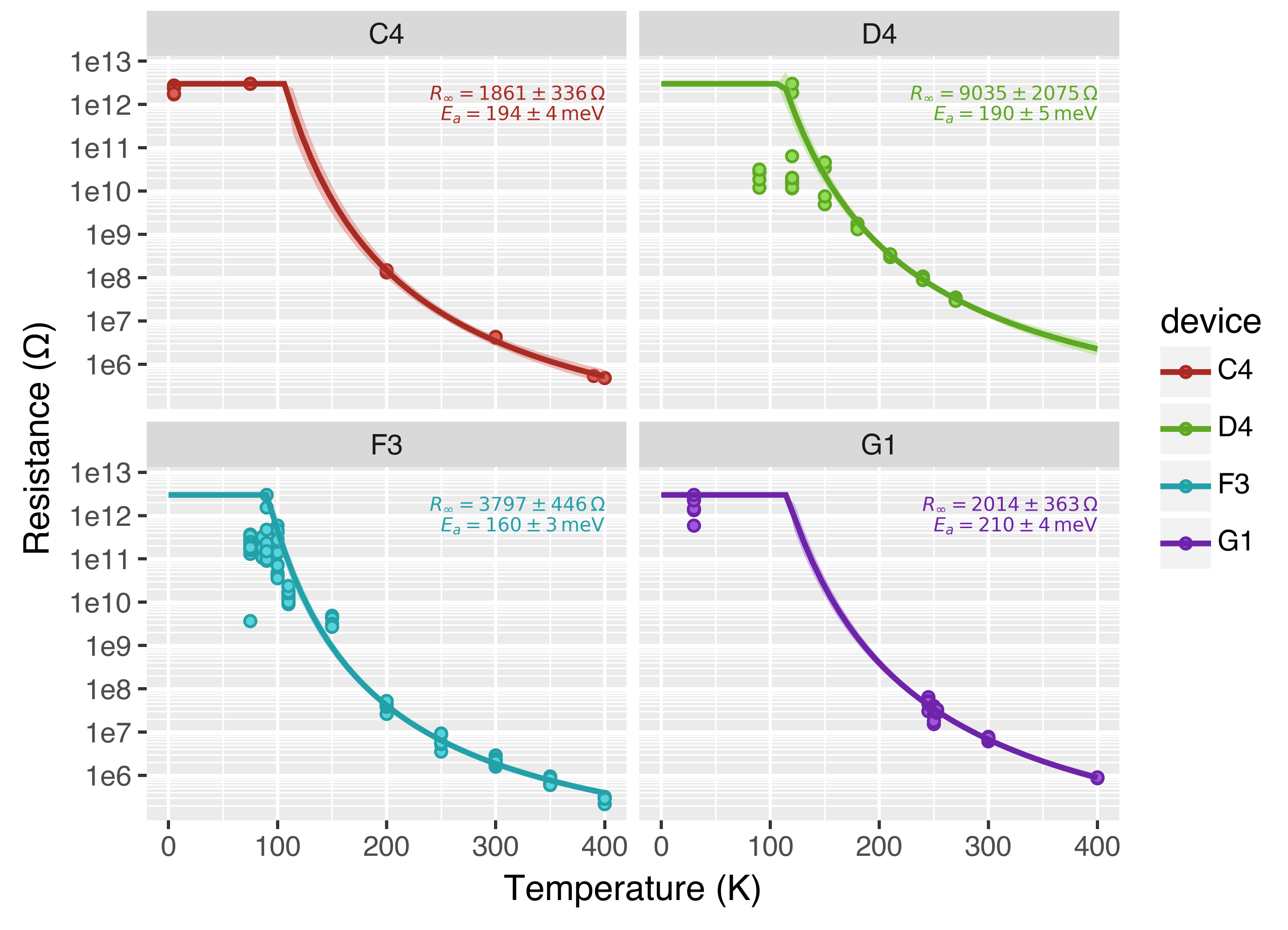}
    \caption{Experimental data and fitting coefficients used in the conductance window model.}
    \label{fig:R0alpha-Ea}
\end{figure}

\noindent The conductance window, or memory window, is defined as the ratio $G_{max} / G_{min}$. A large memory window generally means that the peripheral circuitry will have an easier job in distinguishing the memory states of the device. Empirical models for both $G_{min}$ and $G_{max}$ can be obtained relatively easily by fitting the minimum SET resistance and maximum RESET resistance as a function of temperature. From the known electrical transport laws, the crystalline SET state behavior can be captured by a linear regression, while the amorphous RESET state has been simplified to just an Arrhenius law with an effective activation energy. Both the mean and standard deviation of the parameters have been obtained, and are shown in Figure \ref{fig:R0alpha-Ea}
The model is equivalent to the following:
\begin{align}
  R_{SET} (T) &= R(0) + \alpha \cdot T \notag \\
  R_{RESET} (T) &= R_\infty \exp\left(\frac{E_a}{k_B T}\right)
  \label{eq:Gwindow}
\end{align}
where $R(0)$, $\alpha$, $R_\infty$, $E_a$ are all assumed to be normally distributed. The finite $G_{min}$ and $G_{max}$ are implemented by drawing a sample of $R_{SET,i j}$ and $R_{RESET,i j}$ for each device $G_{i j}$, then clamping the conductance of all devices in the corresponding range $(1 / R_{RESET,i j}, 1 / R_{SET, i j})$. All devices where $R_{SET}$ happened to be higher than the expected value will then result in a $\tilde{G}_{i j} < G_{i j} $ if $G_{i j} \approx G_{max}$, and each device will only be able to reach $G_{min, i j} = 1 / R_{RESET, i j}$. Thus variance in $R(0)$ and $\alpha$ translates to a variance in $G_{max}$ among devices. This can have a measurable effect on accuracy, since the largest weights will be the most affected. This issue can be avoided by choosing a smaller $G'_{max}$ in the scaling factors, ensuring a larger percentage of devices will be able to reach $G_{i j} \approx G'_{max}$.

\subsection*{Conductance Drift}
\begin{figure}[h]
    \centering
    \includegraphics[width=0.45\linewidth]{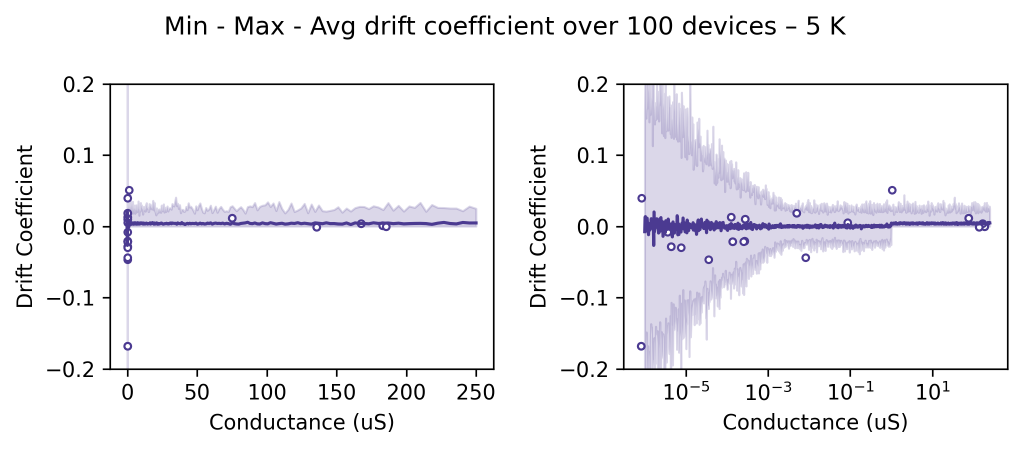}
    \includegraphics[width=0.45\linewidth]{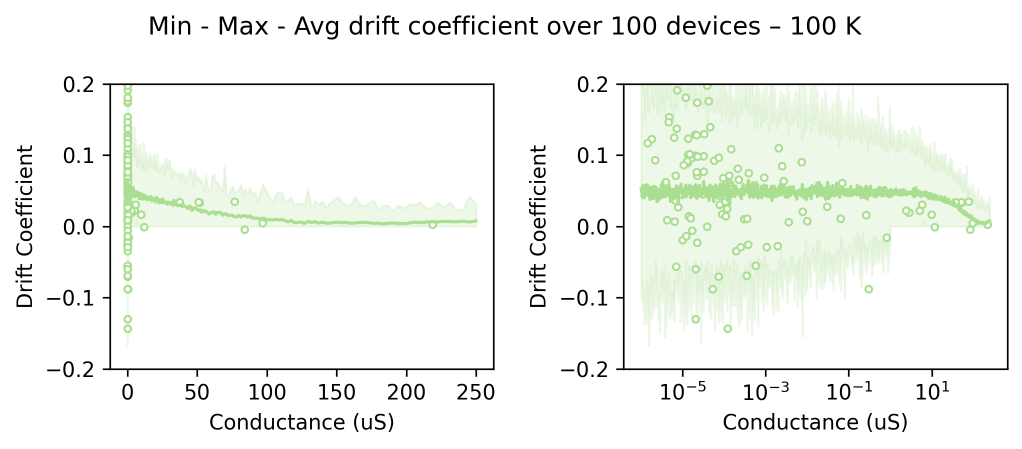}
    \includegraphics[width=0.45\linewidth]{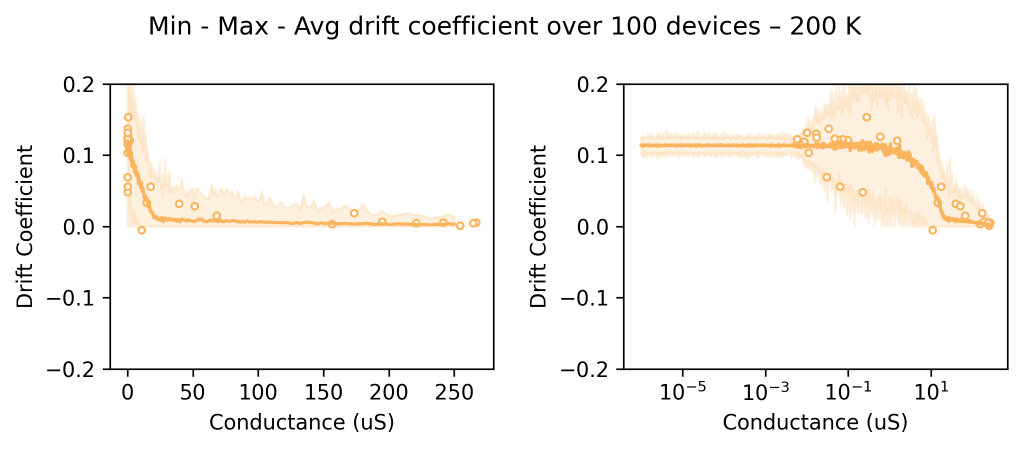}
    \includegraphics[width=0.45\linewidth]{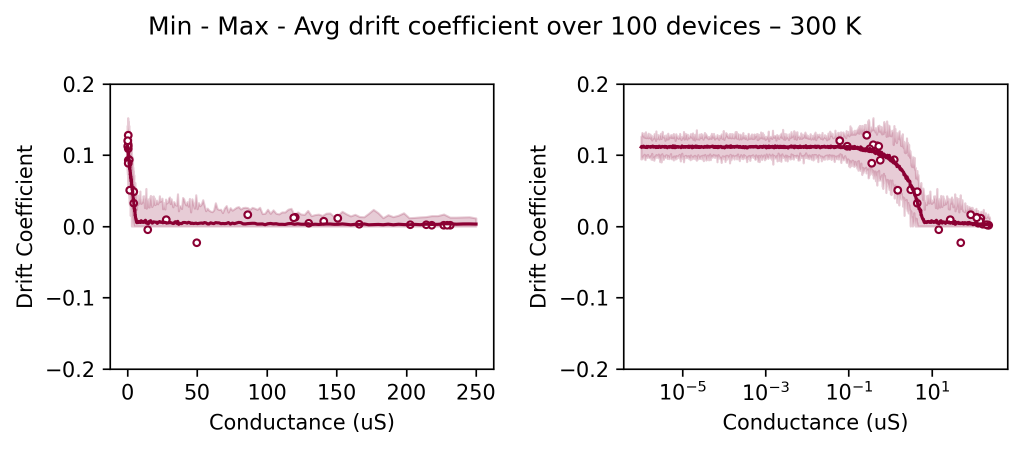}

    \caption{Comparison between real experimental data (dots) and results of the drift empirical model. Solid line shows mean drift coefficient $\nu$ at the given temperature and conductance, shaded area shows average distribution spread.}
    \label{fig:sim-drift}
\end{figure}

\noindent Drift is applied to every device as a power law in time at the read bias:
\[
G(t)=G_0\left(\frac{t}{t_0}\right)^{-\nu(T,G)}
\]
with \(t_0\) the programming time and \(t\) the read time used in simulation. The update acts on the magnitude, \(|G|\), and preserves the sign. For each temperature, \(\nu\) is modeled as a quadratic polynomial of normalized \(G\) with a separate polynomial fit to the absolute residuals versus \(\log_{10}G\) to capture heteroscedasticity. The model returns \(\nu_{\text{mean}}(T,G)\) and \(\nu_{\text{std}}(T,G)\). In simulation a per-device coefficient is drawn as
\[
\nu \sim \mathcal{N}\!\bigl(\nu_{\text{mean}}(T,G),\,\nu_{\text{std}}^2(T,G)\bigr),
\]
then truncated only where required by data consistency:
\[
\text{if } T>100~\text{K} \text{ then } \nu \leftarrow \max(\nu,0),\qquad
\text{if } G>1~\mu\text{S} \text{ then } \nu \leftarrow \max(\nu,0).
\]
Figure~\ref{fig:sim-drift} shows the drift model outputs and compares them with experimental data. We note two details: the extreme width of the memory window at \qty{5}{\kelvin} and \qty{100}{\kelvin} makes the data sparse when observed in the linear regime. Furthermore, as explained in the main text, drift stops following a clear power law at these temperatures. It thus blends into read noise, complicating the analysis.
As such, the observed spread in the distribution is our best guess at estimating the real variance in the drift coefficient.
Finally, the model going beyond the minimum conductance of the experimental data is an artifact of this specific visualization. In the complete simulation this issue never arises since the limited conductance window is applied before calculating the drift coefficients.

\subsection*{Read Noise}
\noindent Read noise is modeled from band-limited PSD fits of current fluctuations. The normalized PSD is
\[
\frac{S_{II}(f)}{I^2}=\frac{Q(T,G)}{f^{\alpha(T,G)}} ,
\]
fitted per temperature as a function of conductance. The empirical models return \(\log_{10}Q(T,G)\) and \(\alpha(T,G)\) from the 0.01–1\,Hz dataset. In simulation we convert to variance of conductance fluctuations by integrating the PSD over the read bandwidth:
\[
\sigma_G^2(T,G;t,\tau_\text{read})
= G^2 \cdot\frac{Q(T,G)}{1-\alpha(T,G)}\cdot\left[\left(f_\text{max}\right)^{\,1-\alpha}-\left(f_\text{min}\right)^{\,1-\alpha}\right],
\]
with \(f_\text{min}=1/t\) (elapsed time since programming) and \(f_\text{max}=1/(2\,\tau_\text{read})\) (Nyquist of the read duration). For \(\alpha\to1\) the limit \(\sigma_G^2=G^2\,Q\,\ln(f_\text{max}/f_\text{min})\) is used numerically by setting \(\alpha=1.0001\).

\begin{figure}[h]
    \centering
    \includegraphics[width=1\linewidth]{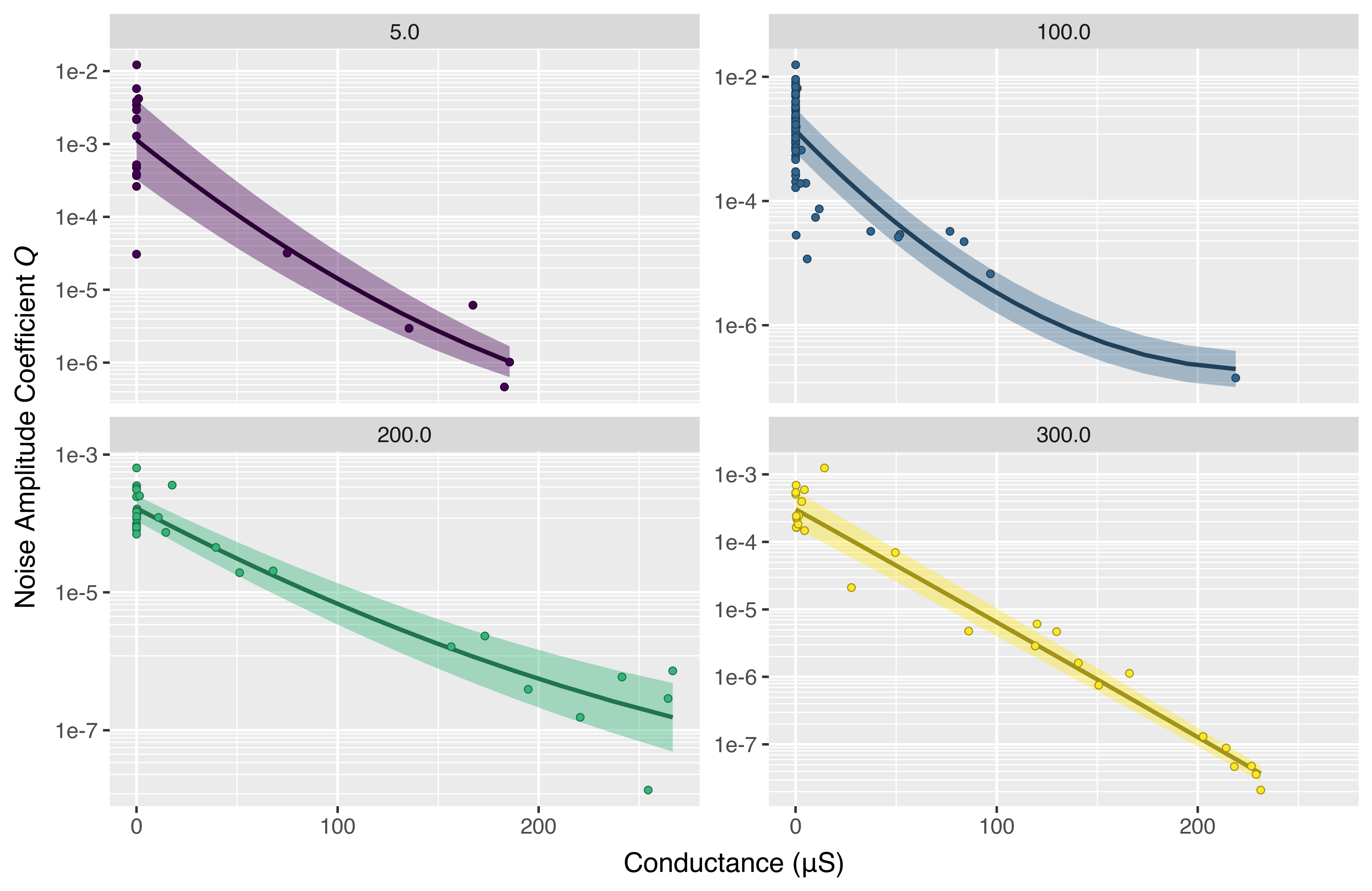}
    \includegraphics[width=1\linewidth]{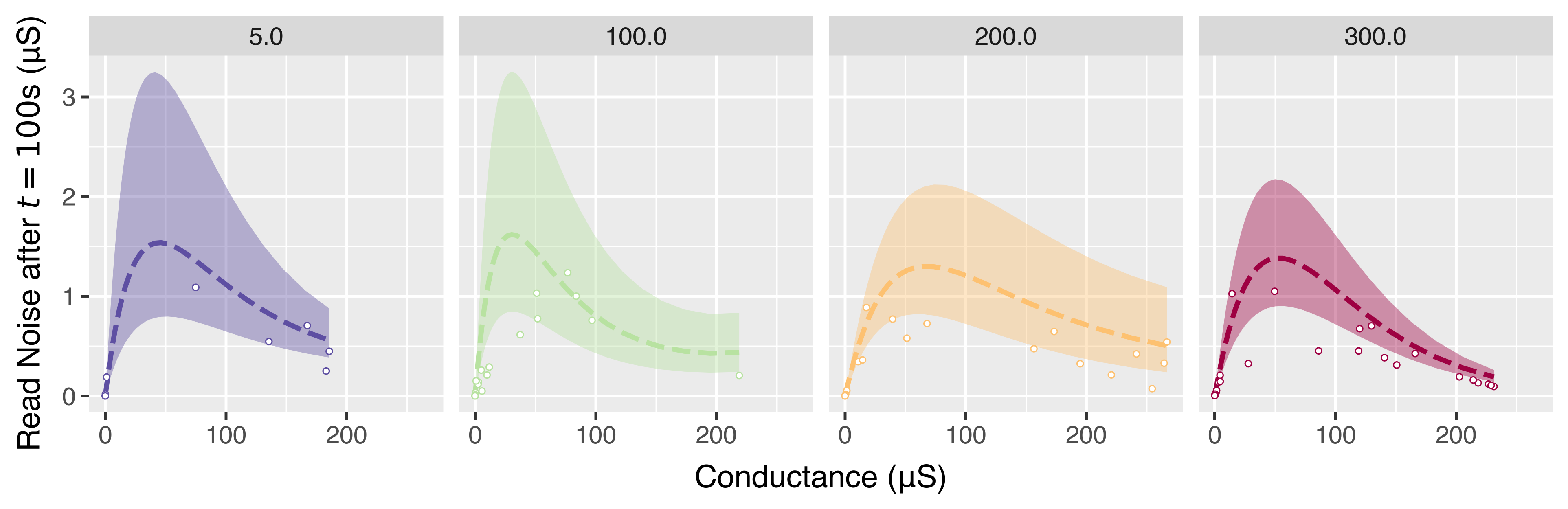}
    \caption{
    Top: Comparison between experimental data (dots) and trend fit for the read noise coefficient amplitude. Bottom:
    Comparison between PSD integrated for $\qty{0.01}{\hertz}<f<\qty{10}{\hertz}$ (dots) and outputs of the read noise empirical model (dashed line + shaded region).
    The line shows the average standard deviation from target, while the shaded area shows the spread of the generated distribution.
    }
    \label{fig:sim-readnoise}
\end{figure}

\noindent The noise injection is additive and elementwise on signed conductance:
\[
G \leftarrow G + \mathcal{N}\!\bigl(0,\,\sigma_G^2(T,|G|;t,\tau_\text{read})\bigr),
\]
with \(Q=10^{\log_{10}Q(T,|G|)}\) and \(\alpha=\alpha(T,|G|)\).
Figure~\ref{fig:sim-readnoise} shows the read noise model outputs and compares them with experimental data.
The model seems to often overestimate the amount of read noise, but we have found the deviation to have a negligible effect on the simulation results. The deviation most likely comes from deviations between the fitted $1/f^\alpha$ PSD and the real PSD caused by random telegraph noise (RTN) and similar phenomena.

\subsection*{Array Programming}
\noindent Although our experiments did not run an iterative write-verify programming loop, the reproducibility of the curves in Figure~2a (main text) confirms that such a procedure should still work at any temperature. A possible problem is that, since the conductance window (defined as $G_\text{max} / G_\text{min}$) grows above $10^7$ below $\SI{100}{\kelvin}$, a conventional verify step would need an ADC fine enough to resolve that span, requiring $\sim24\,$bits of accuracy. This obstacle can be removed with gradient-based crossbar programming\cite{JBuchelASebastian202212}, which updates weights collectively and never has to read a single device with high precision.

\clearpage

\begin{table}[h]
    \sisetup{
      round-mode        = figures ,   
      round-precision   = 3      ,   
      scientific-notation = fixed ,   
      exponent-product  = \cdot ,     
      group-digits = none,
      group-minimum-digits = 4,
      table-number-alignment = center 
    }
    \input{supplementary/simulation_metrics}
    \caption{Input and output metrics for the simulation of PCM-based MVM.}
    \label{tab:metrics}
\end{table}

\clearpage
\def\url#1{}
\bibliographystyle{naturemag}
\bibliography{ReferencesFull}

%% file: supplementary/simulation_metrics.tex
\begin{tabular}{l S S S S S}
    \toprule
    {} & {} & {$\ell_2\,\text{error}\,(\%)$} & {$G_{avg}\,(\mu\text{S})$} & {$I_{avg}\,(\mu\text{A})$} & {$P_{diss}\,(\mu\text{W})$} \\
    {$T\,(\text{K})$} & {$G_{max}\,(\mu\text{S})$} & {} & {} & {} & {} \\
    \midrule
    \multirow[c]{9}{*}{300} & 147.772114 & 11.187989 & 22.657956 & 20.861605 & 4867.069087 \\
     & 118.217691 & 10.971671 & 18.071001 & 16.763831 & 3883.142445 \\
     & 73.886057 & 12.359294 & 11.103706 & 10.480904 & 2387.043098 \\
     & 44.331634 & 15.119448 & 6.378265 & 6.173255 & 1371.612816 \\
     & 8.940213 & 51.973583 & 0.717268 & 0.773430 & 153.711308 \\
     & 1.802820 & 100.911378 & 0.186008 & 0.253139 & 39.648497 \\
     & 0.369430 & 442.956110 & 0.147986 & 0.237945 & 31.391959 \\
     & 0.073886 & 2302.448774 & 0.146400 & 0.237646 & 31.041051 \\
     & 0.014777 & 11638.738318 & 0.146387 & 0.237639 & 31.038036 \\
    \midrule
    \multirow[c]{9}{*}{200} & 156.605984 & 17.870122 & 22.101403 & 21.299413 & 4749.390557 \\
     & 125.284787 & 20.000595 & 17.146038 & 16.816187 & 3683.885890 \\
     & 78.302992 & 28.184538 & 9.581483 & 9.821453 & 2054.000819 \\
     & 46.981795 & 41.920126 & 4.637300 & 5.069156 & 991.982376 \\
     & 9.474662 & 66.636573 & 0.540991 & 0.548444 & 115.976454 \\
     & 1.910593 & 71.071453 & 0.096074 & 0.094097 & 20.608437 \\
     & 0.391515 & 72.216294 & 0.019846 & 0.019283 & 4.256595 \\
     & 0.078303 & 88.324697 & 0.006139 & 0.008084 & 1.308008 \\
     & 0.015661 & 321.096109 & 0.004299 & 0.007446 & 0.908789 \\
    \midrule
    \multirow[c]{9}{*}{100} & 166.563191 & 28.406561 & 20.306142 & 19.581975 & 4355.055840 \\
     & 133.250552 & 31.180908 & 15.663559 & 15.076225 & 3359.562053 \\
     & 83.281595 & 36.085570 & 9.191353 & 8.780069 & 1971.426942 \\
     & 49.968957 & 39.832314 & 5.267597 & 5.006353 & 1129.817139 \\
     & 10.077073 & 45.433249 & 1.009952 & 0.959693 & 216.555596 \\
     & 2.032071 & 47.706154 & 0.207680 & 0.202206 & 44.493808 \\
     & 0.416408 & 49.825430 & 0.043495 & 0.043306 & 9.316620 \\
     & 0.083282 & 52.009355 & 0.008909 & 0.009083 & 1.907762 \\
     & 0.016656 & 54.393071 & 0.001825 & 0.001902 & 0.390676 \\
    \midrule
    \multirow[c]{9}{*}{5} & 177.270733 & 9.697866 & 27.732993 & 25.492125 & 5958.991225 \\
     & 141.816587 & 10.459120 & 22.186126 & 20.404045 & 4766.808142 \\
     & 88.635367 & 12.318972 & 13.865244 & 12.768818 & 2978.623591 \\
     & 53.181220 & 14.412811 & 8.319614 & 7.673392 & 1787.125843 \\
     & 10.724879 & 18.587278 & 1.685998 & 1.554727 & 362.093618 \\
     & 2.162703 & 21.760860 & 0.354319 & 0.325422 & 76.066181 \\
     & 0.443177 & 22.648896 & 0.072891 & 0.067222 & 15.650866 \\
     & 0.088635 & 22.694558 & 0.014579 & 0.013446 & 3.130288 \\
     & 0.017727 & 22.807786 & 0.002919 & 0.002690 & 0.626770 \\
    \bottomrule
\end{tabular}